\documentclass[aps,prd,showpacs,twocolumn,superscriptaddress]{revtex4-1}
\usepackage{longtable}
\usepackage{amsmath}
\usepackage{graphicx}
\usepackage{dcolumn}

\newcommand{\mm}{M_{\rm miss}}
\newcommand{\de}{\Delta E}
\newcommand{\up}{\Upsilon}
\newcommand{\upns}{\up(nS)}
\newcommand{\upf}{\up(10860)}
\newcommand{\pipi}{\pi^+\pi^-}
\newcommand{\pnpn}{\pi^0\pi^0}

\newcommand{\br}{{\cal B}}

\newcommand{\ee}{e^+e^-}
\newcommand{\mumu}{\mu^+\mu^-}
\newcommand{\lplm}{\ell^+\ell^-}
\newcommand{\mev}{{\rm MeV}}
\newcommand{\mevcc}{{\rm MeV}/c^2}
\newcommand{\gevcc}{{\rm GeV}/c^2}
\newcommand{\zb}{Z_b}
\newcommand{\zbn}{Z_b^0}
\newcommand{\zbc}{Z_b^\pm}
\newcommand{\zbnf}{\zbn(10610)}
\newcommand{\zbns}{\zbn(10650)}
\newcommand{\fix}{(fixed)}
\newcommand{\tl}{2\ln {\mathcal L}}
\newcommand{\dtl}{\Delta(-\tl)}
\newcommand{\anr}{a^{\rm nr}}
\newcommand{\phinr}{\phi^{\rm nr}}
\newcommand{\fsig}{f_{\rm sig}}
\newcommand{\sborn}{\sigma_{\rm Born}}
\newcommand{\svis}{\sigma_{\rm vis}}
\newcommand{\sbb}{\sigma_{b{\bar b}}}
\newcommand{\seeuppp}{\sigma(\ee\to\upns\pnpn)}
\newcommand{\disr}{(1+\delta_{\rm ISR})}

\begin{document}

\title{\boldmath First Observation of the $\zbnf$ in a Dalitz Analysis of 
$\upf\to\upns\pnpn$}


\begin{abstract}
  We report the first observation of $\upf\to\up(1,2,3S)\pnpn$ decays.
  The neutral partner of the $\zbc(10610)$, the $\zbnf$ decaying to 
  $\up(2,3S)\pi^0$, is observed for the first time with a $6.5\sigma$ 
  significance using a Dalitz analysis of $\upf\to\up(2,3S)\pnpn$ decays.
  The results are obtained with a $121.4\,{\rm fb}^{-1}$ data sample 
  collected with the Belle detector at the $\upf$ resonance at the 
  KEKB asymmetric-energy $\ee$ collider.
\end{abstract}

\pacs{14.40.Pq, 13.25.Gv, 12.39.Pn}

\noaffiliation
\affiliation{University of the Basque Country UPV/EHU, 48080 Bilbao}
\affiliation{University of Bonn, 53115 Bonn}
\affiliation{Budker Institute of Nuclear Physics SB RAS and Novosibirsk State University, Novosibirsk 630090}
\affiliation{Faculty of Mathematics and Physics, Charles University, 121 16 Prague}
\affiliation{Chiba University, Chiba 263-8522}
\affiliation{University of Cincinnati, Cincinnati, Ohio 45221}
\affiliation{Deutsches Elektronen--Synchrotron, 22607 Hamburg}
\affiliation{Justus-Liebig-Universit\"at Gie\ss{}en, 35392 Gie\ss{}en}
\affiliation{II. Physikalisches Institut, Georg-August-Universit\"at G\"ottingen, 37073 G\"ottingen}
\affiliation{Gyeongsang National University, Chinju 660-701}
\affiliation{Hanyang University, Seoul 133-791}
\affiliation{University of Hawaii, Honolulu, Hawaii 96822}
\affiliation{High Energy Accelerator Research Organization (KEK), Tsukuba 305-0801}
\affiliation{Hiroshima Institute of Technology, Hiroshima 731-5193}
\affiliation{Ikerbasque, 48011 Bilbao}
\affiliation{Indian Institute of Technology Guwahati, Assam 781039}
\affiliation{Indian Institute of Technology Madras, Chennai 600036}
\affiliation{Institute of High Energy Physics, Chinese Academy of Sciences, Beijing 100049}
\affiliation{Institute of High Energy Physics, Vienna 1050}
\affiliation{Institute for High Energy Physics, Protvino 142281}
\affiliation{INFN - Sezione di Torino, 10125 Torino}
\affiliation{Institute for Theoretical and Experimental Physics, Moscow 117218}
\affiliation{J. Stefan Institute, 1000 Ljubljana}
\affiliation{Kanagawa University, Yokohama 221-8686}
\affiliation{Institut f\"ur Experimentelle Kernphysik, Karlsruher Institut f\"ur Technologie, 76131 Karlsruhe}
\affiliation{Korea Institute of Science and Technology Information, Daejeon 305-806}
\affiliation{Korea University, Seoul 136-713}
\affiliation{Kyungpook National University, Daegu 702-701}
\affiliation{\'Ecole Polytechnique F\'ed\'erale de Lausanne (EPFL), Lausanne 1015}
\affiliation{Faculty of Mathematics and Physics, University of Ljubljana, 1000 Ljubljana}
\affiliation{Luther College, Decorah, Iowa 52101}
\affiliation{University of Maribor, 2000 Maribor}
\affiliation{Max-Planck-Institut f\"ur Physik, 80805 M\"unchen}
\affiliation{School of Physics, University of Melbourne, Victoria 3010}
\affiliation{Moscow Physical Engineering Institute, Moscow 115409}
\affiliation{Moscow Institute of Physics and Technology, Moscow Region 141700}
\affiliation{Graduate School of Science, Nagoya University, Nagoya 464-8602}
\affiliation{Kobayashi-Maskawa Institute, Nagoya University, Nagoya 464-8602}
\affiliation{Nara Women's University, Nara 630-8506}
\affiliation{National Central University, Chung-li 32054}
\affiliation{National United University, Miao Li 36003}
\affiliation{Department of Physics, National Taiwan University, Taipei 10617}
\affiliation{H. Niewodniczanski Institute of Nuclear Physics, Krakow 31-342}
\affiliation{Nippon Dental University, Niigata 951-8580}
\affiliation{Niigata University, Niigata 950-2181}
\affiliation{University of Nova Gorica, 5000 Nova Gorica}
\affiliation{Osaka City University, Osaka 558-8585}
\affiliation{Pacific Northwest National Laboratory, Richland, Washington 99352}
\affiliation{Panjab University, Chandigarh 160014}
\affiliation{University of Pittsburgh, Pittsburgh, Pennsylvania 15260}
\affiliation{Research Center for Electron Photon Science, Tohoku University, Sendai 980-8578}
\affiliation{University of Science and Technology of China, Hefei 230026}
\affiliation{Seoul National University, Seoul 151-742}
\affiliation{Soongsil University, Seoul 156-743}
\affiliation{Sungkyunkwan University, Suwon 440-746}
\affiliation{School of Physics, University of Sydney, NSW 2006}
\affiliation{Tata Institute of Fundamental Research, Mumbai 400005}
\affiliation{Excellence Cluster Universe, Technische Universit\"at M\"unchen, 85748 Garching}
\affiliation{Toho University, Funabashi 274-8510}
\affiliation{Tohoku Gakuin University, Tagajo 985-8537}
\affiliation{Tohoku University, Sendai 980-8578}
\affiliation{Department of Physics, University of Tokyo, Tokyo 113-0033}
\affiliation{Tokyo Institute of Technology, Tokyo 152-8550}
\affiliation{Tokyo Metropolitan University, Tokyo 192-0397}
\affiliation{Tokyo University of Agriculture and Technology, Tokyo 184-8588}
\affiliation{University of Torino, 10124 Torino}
\affiliation{CNP, Virginia Polytechnic Institute and State University, Blacksburg, Virginia 24061}
\affiliation{Wayne State University, Detroit, Michigan 48202}
\affiliation{Yamagata University, Yamagata 990-8560}
\affiliation{Yonsei University, Seoul 120-749}

\author{P.~Krokovny}\affiliation{Budker Institute of Nuclear Physics SB RAS and Novosibirsk State University, Novosibirsk 630090} 
\author{A.~Bondar}\affiliation{Budker Institute of Nuclear Physics SB RAS and Novosibirsk State University, Novosibirsk 630090} 

  \author{I.~Adachi}\affiliation{High Energy Accelerator Research Organization (KEK), Tsukuba 305-0801} 
  \author{H.~Aihara}\affiliation{Department of Physics, University of Tokyo, Tokyo 113-0033} 
  \author{K.~Arinstein}\affiliation{Budker Institute of Nuclear Physics SB RAS and Novosibirsk State University, Novosibirsk 630090} 
  \author{D.~M.~Asner}\affiliation{Pacific Northwest National Laboratory, Richland, Washington 99352} 
  \author{V.~Aulchenko}\affiliation{Budker Institute of Nuclear Physics SB RAS and Novosibirsk State University, Novosibirsk 630090} 
  \author{T.~Aushev}\affiliation{Institute for Theoretical and Experimental Physics, Moscow 117218} 
  \author{T.~Aziz}\affiliation{Tata Institute of Fundamental Research, Mumbai 400005} 
  \author{A.~M.~Bakich}\affiliation{School of Physics, University of Sydney, NSW 2006} 
% ...
\collaboration{The Belle Collaboration}


\maketitle


\section{Introduction}
Two charged bottomonium-like resonances, $\zbc(10610)$ and $\zbc(10650)$, 
have been observed by the Belle Collaboration~\cite{zb_paper} in the 
$\upns\pi^\pm$ invariant mass in $\upf\to\upns\pipi$ decays ($n=1,2,3$) and in 
$h_b(mP)\pi^\pm$ mass spectra in the recently observed $\upf\to h_b(mP)\pipi$ 
decays ($m=1,2$)~\cite{hb_paper}.
An angular analysis suggests that these states have 
$I^G(J^P)=1^+(1^+)$ quantum numbers~\cite{zb_helicity}.
Analysis of the quark composition of the initial and final states 
allows us to assert that these hadronic objects are the first examples of 
states of an exotic nature with a $b\bar{b}$ quark pair:
$\zb$ should be comprised of (at least) four quarks. Several models have
been proposed to describe the internal structure of these 
states~\cite{danilkin, bugg, karliner}.
The proximity of the $\zbc(10610)$ and $\zbc(10650)$ masses to 
thresholds for the open beauty channels $B^*\bar{B}$ and $B^*\bar{B}^*$  
suggests a ``molecular'' structure for these states, 
which is consistent with many of their observed properties~\cite{zb_molecular}.
More recently, Belle reported the observation of both
$\zbc(10610)$ and $\zbc(10650)$ in an analysis of the three-body
$\upf\to [B^{(*)}B^*]^\mp\pi^\pm$ decay~\cite{y5s_3body}. The dominant $\zb$ 
decay mode is found to be $B^{(*)}B^*$, supporting the molecular hypothesis.
It would be natural to expect the existence of neutral partners of 
these states. This motivates us to search for $\zbn$ in the resonant 
substructure of $\upf\to\upns\pnpn$ decays.

\section{Data sample and Detector}
We use a $(121.4\pm 1.7)\ {\rm fb}^{-1}$ data sample collected on the peak
of the $\upf$ resonance with the Belle detector~\cite{belle} at the KEKB 
asymmetric-energy $\ee$ collider~\cite{kekb}.
The Belle detector is a large-solid-angle magnetic spectrometer that
consists of a silicon vertex detector, a central drift chamber (CDC), an array
of aerogel threshold Cherenkov counters (ACC), a barrel-like arrangement of
time-of-flight scintillation counters, and an electromagnetic calorimeter (ECL)
comprised of CsI(Tl) crystals located inside a superconducting solenoid
that provides a 1.5~T magnetic field. An iron flux return located outside
the coil is instrumented to detect $K_L^0$ mesons and to identify muons (KLM).
The detector is described in detail elsewhere~\cite{belle}.

\section{Signal Selection}
$\upf$ candidates are formed from $\upns\pnpn$ ($n=1,2,3$) combinations.
We reconstruct $\upns$ candidates from pairs of leptons ($\ee$ and $\mumu$,
referred to as $\lplm$) with an invariant mass between 8 and 11~$\gevcc$.
An additional decay channel is used for the $\up(2S)$: 
$\up(2S)\to\up(1S)[\lplm]\pipi$.
Charged tracks are required to have a transverse momentum, $p_t$,
greater than $50$~MeV$/c$. 
We also impose a requirement on the impact parameters of
$dr<0.3$~cm and $|dz|<2.0$~cm, where $dr$ and $dz$ are the impact 
parameters in the $r$-$\phi$ and longitudinal directions, respectively.
Muon candidates are required to have associated hits in the KLM detector 
that agree with the extrapolated trajectory of a charged track provided by 
the drift chamber~\cite{muid}. Electron candidates are identified 
based on the ratio of ECL shower energy to the track momentum, ECL shower 
shape, $dE/dx$ from the CDC, and the ACC response~\cite{elid}.
No particle identification requirement is imposed for the pions.
Candidate $\pi^0$ mesons are selected from pairs of photons with an 
invariant mass within 15~$\mevcc$ ($3\sigma$) of the nominal $\pi^0$ mass.
An energy greater than 50 (75) MeV is required for each photon in the 
barrel (endcap).
We use the quality of the $\pi^0$ mass-constrained fits,
$\chi^2(\pi^0_1)+\chi^2(\pi^0_2)$, to suppress the background.
This sum must be less than 20 (10) for the $\upns\to\mumu$, 
$\up(1S)\pipi$ ($\upns\to\ee$).

\begin{figure*}
  \includegraphics[width=0.32\textwidth] {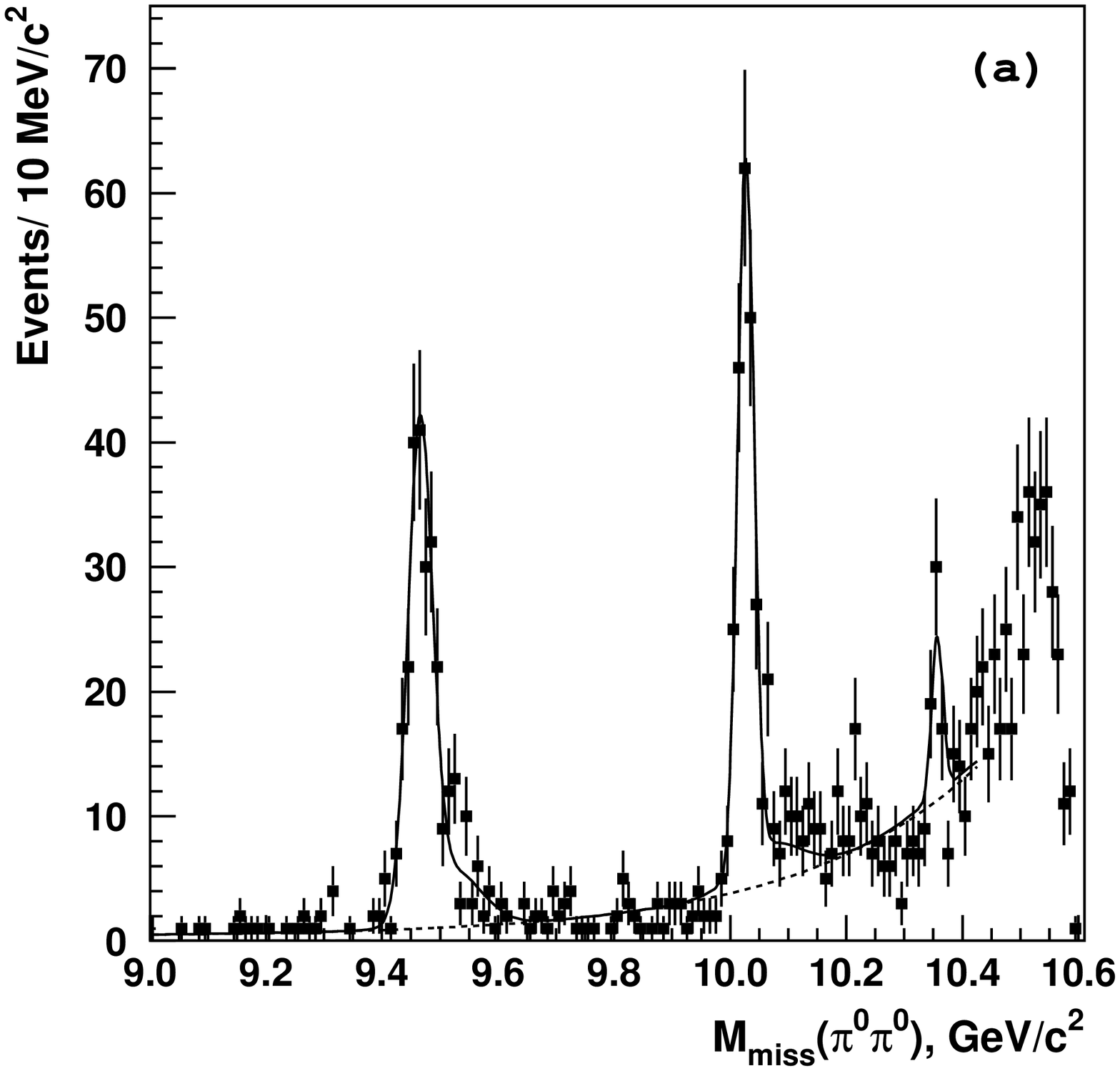}
  \includegraphics[width=0.32\textwidth] {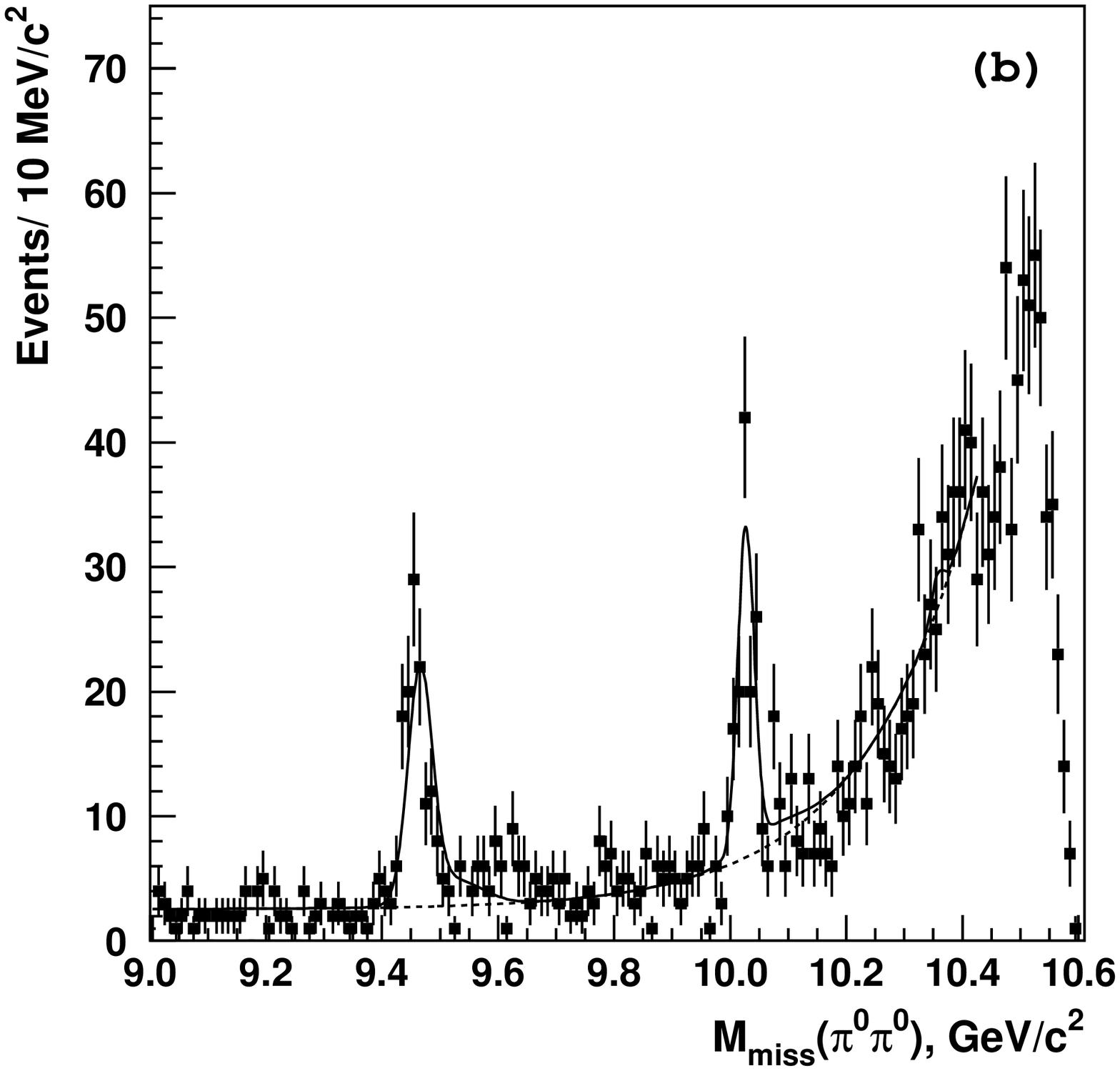}
  \includegraphics[width=0.32\textwidth] {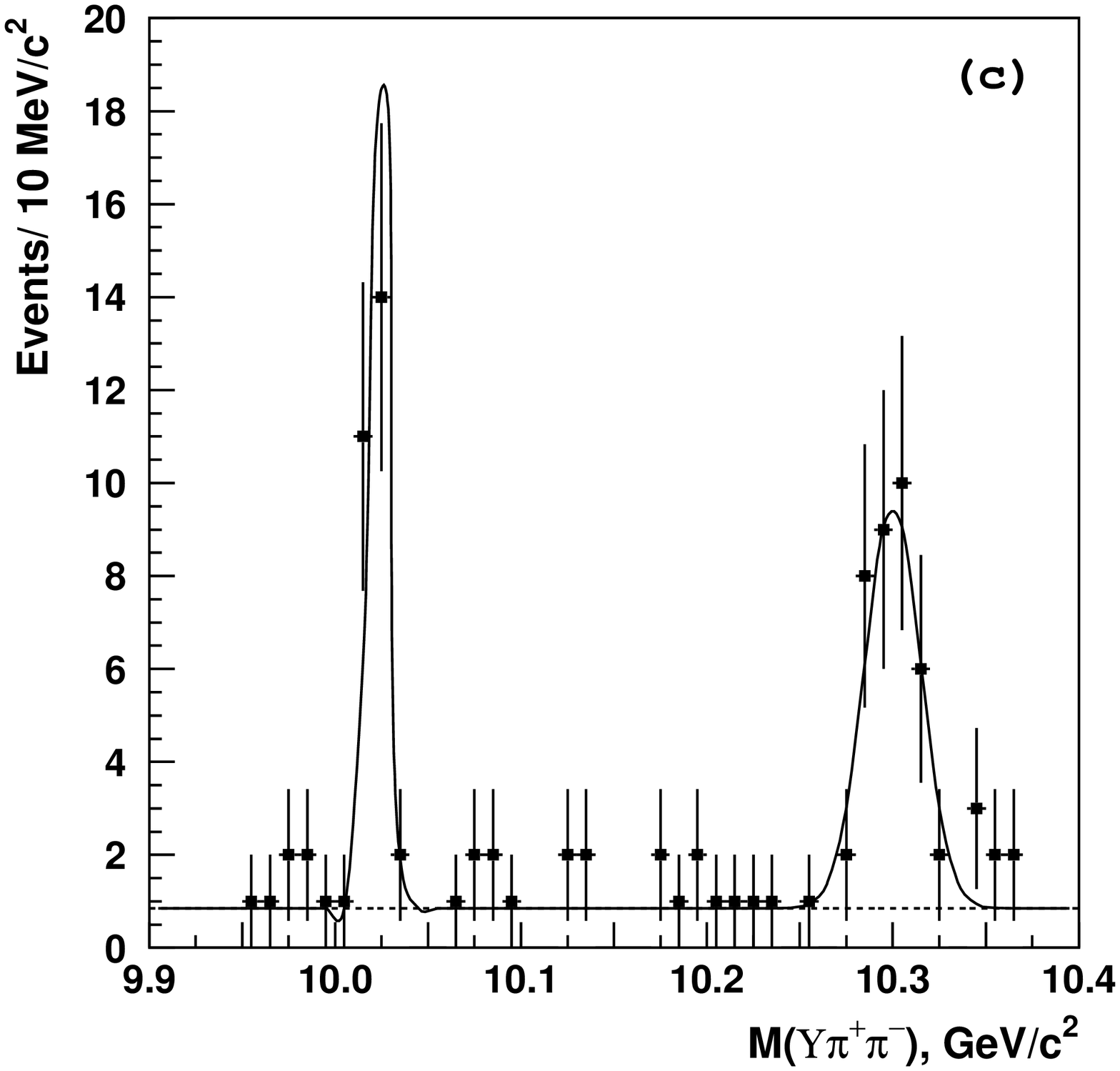}
  \caption{The $\pnpn$ missing mass distribution for $\upns\pnpn$ 
    candidates, using (a) $\upns\to\mumu$ and (b) $\upns\to\ee$ candidates.
    The $M(\up(1S)\pipi)$ distribution for $\up(2S)\to\up(1S)\pipi$ 
    candidates is shown in (c). Histograms represent the data. 
    In each panel, the solid curve shows the fit result 
    while the dashed curve corresponds to the background contribution.}
  \label{fig:mmpnpn}
\end{figure*}

We use the energy difference $\de= E_{\rm cand}-E_{\rm CM}$ and momentum 
$P$ to suppress background, where $E_{\rm cand}$ and $P$ are the
energy and momentum of the reconstructed $\upf$ candidate in the 
center-of-mass (c.m.) frame, and $E_{\rm CM}$ is the c.m. energy of the 
two beams. 
$\upf$ candidates must satisfy the requirements
$-0.2$~GeV$<\de<0.14$~GeV and $P<0.2$~GeV$/c$.
The potentially large background from QED processes such as 
$\ee\to\lplm (n)\gamma$ is suppressed using the missing mass associated 
with the $\lplm$ system, 
calculated as $\mm(\lplm)=\sqrt{(E_{\rm CM}-E_{\lplm})^2-P_{\lplm}^2}$,
where $E_{\lplm}$ and $P_{\lplm}$ are the energy and momentum of the 
$\lplm$ system measured in the c.m. frame. We require 
$\mm(\lplm)>0.15\,(0.30)$~$\gevcc$ for the $\upns\to\mumu$ ($\ee$).
We select the candidate with the smallest $\chi^2(\pi^0_1)+\chi^2(\pi^0_2)$ 
in the rare cases (1-2\%) when there is more than one candidate 
in the event.
Figures~\ref{fig:mmpnpn}~(a) and (b) show the $\mm(\pnpn)$ distributions 
for the $\upf\to\upns[\lplm]\pnpn$ candidates, which are evaluated similarly 
to $\mm(\lplm)$. Clear peaks of the $\up(1S)$, $\up(2S)$ and $\up(3S)$ 
can be seen.

For $\upf\to\up(2S)[\up(1S)\pipi]\pnpn$ decays, $\up(1S)$ candidates 
are selected from $\lplm$ pairs with invariant mass within 
$150\,\mevcc$ of the nominal $\up(1S)$ mass.
A mass-constrained fit is used for $\up(1S)$ candidates to improve 
the momentum resolution. We apply the requirements on $\de$ and $P$ 
for $\upf$ candidates described earlier. 
We select signal candidates with the invariant mass of $\up(1S)\pipi$
within $20\,\mevcc$ of the nominal $\up(2S)$ mass.
Figure~\ref{fig:mmpnpn}~(c) shows the $M(\up(1S)\pipi)$ distribution for the 
$[\up(1S)\pipi]\pnpn$ events. The clear peak of the $\up(2S)$ can be seen.
The peak around $10.3\,\gevcc$ corresponds to a
reflection from the decay $\upf\to\up(2S)\pipi$, $\up(2S)\to\up(1S)\pnpn$.

\section{\boldmath $\ee\to\upns\pnpn$ cross sections at $\upf$}
The signal yields for $\upf\to\upns[\lplm]\pnpn$ decays are extracted 
by a binned maximum likelihood fit to the $\mm(\pnpn)$ distributions.
The signal probability density function (PDF) is described by a sum of two 
Gaussians for each $\upns$ resonance with parameters fixed from the signal 
Monte Carlo (MC) sample. 
The correctly reconstructed events ($\sim 80\%$) are described by a
core Gaussian with the resolution of 
21, 14 and 10 $\mevcc$ for $\up(1S)$, $\up(2S)$ and $\up(3S)$, respectively.
A sizable fraction ($\sim 20\%$) of events with misrecostructed $\gamma$
from $\pi^0$ decay are described by a wider Gaussian with a shifted mean.
The background PDF is 
parameterized by the sum of a constant and an exponential function.

For the $\up(2S)[\up(1S)\pipi]$ decay, we fit the invariant mass
of $\up(1S)\pipi$.
The signal PDF is described by a Gaussian function with a resolution of 
$5\, \mevcc$ (fixed from signal MC). The background PDF is described by a 
constant. The cross-feed from the decay $\upf\to\up(2S)[\up(1S)\pnpn]\pipi$ 
contributes as a broad peak around $10.3$~$\gevcc$. Its shape is 
parameterized by a Gaussian function with parameters fixed from MC.
The fit results are also shown in Fig.~\ref{fig:mmpnpn}~(a)-(c).

Though $\upns\pnpn$ final states are expected to be produced from
the decay of the $\upf$ resonance, here we present the signal rates
as the cross sections of $\ee\to\upns\pnpn$ since the fraction of 
the resonance among $b\bar{b}$ hadronic events is unknown and the
energy dependence of the $\upns\pipi$ yield is found to be rather different
from that of $b\bar{b}$ hadronic events~\cite{br_ypipi}. 
Table~\ref{tab:yield} summarizes the signal yield, MC efficiency and measured 
visible cross section (with only the statistical uncertainty shown). 
The reconstruction efficiency is obtained from MC using the matrix element
determined from the Dalitz plot fit described below. The systematic 
uncertainty due to the corresponding fit model is found to be negligible.
The visible cross section is calculated from 
\begin{equation}
\svis=\frac{N_{\rm sig}}{\epsilon\, \br(\upns\to X)\, {\cal L}}\, ,
\end{equation}
where $N_{\rm sig}$ is the number of signal events, $\epsilon$ is the 
reconstruction efficiency, $\br(\upns\to X)$ is the  
branching fraction of the $\upns$ to the reconstructed final state 
$X$ ($\mumu$, $\ee$ or $\up(1S)[\lplm]\pipi$),
and ${\cal L}$ is the integrated luminosity.
The cross section corrected for the initial state radiation (ISR),
the ``dressed'' cross section, is calculated as
\begin{equation}
\sigma = \svis / \disr\, .
\end{equation}
The initial state radiation (ISR) correction factor,
$\disr=0.666\pm 0.013$, is determined using the formulae in 
Ref.~\cite{kuraev}. We assume the energy dependence of $\ee\to\upns\pnpn$
to be the same as for the isospin-related channel $\ee\to\upns\pipi$, 
given by Ref.~\cite{br_ypipi}.
Since  $\br(\up(3S)\to\ee)$ has not been measured, we assume it to be equal to
$\br(\up(3S)\to\mumu)$.

\begin{table*}
\caption{Signal yield ($N_{\rm sig}$), MC efficiency, visible cross section
  ($\svis$), definition of the signal region, number of selected events and 
  fraction of signal events ($\fsig$).}

\medskip
\label{tab:yield}
\begin{tabular*}{\textwidth}{@{\extracolsep{\fill}}lcccccc}\hline\hline
Final state & $N_{\rm sig}$ & $\epsilon$, \% &$\svis$, pb & Signal region, $\gevcc$ & Events & $\fsig$\\ 
\hline
$\up(1S)\to\mumu$ & $261\pm 15$ & $11.2$ & $0.77\pm 0.04$ & $9.41<\mm(\pnpn)<9.53$  & 247 & $0.95$\\
$\up(1S)\to\ee$   & $123\pm 13$ & $5.61$ & $0.76\pm 0.08$ & $9.41<\mm(\pnpn)<9.53$  & 140 & $0.78$\\
$\up(2S)\to\mumu$ & $241\pm 18$ & $8.04$ & $1.28\pm 0.10$ & $9.99<\mm(\pnpn)<10.07$ & 253 & $0.87$\\
$\up(2S)\to\ee$   & $108\pm 13$ & $3.58$ & $1.30\pm 0.16$ & $9.99<\mm(\pnpn)<10.07$ & 151 & $0.66$\\
$\up(2S)\to\up(1S)\pipi$&$24\pm 5$&$2.27$& $1.00\pm 0.21$ &$10.00<M(\up\pipi)<10.05$&  28 & $0.86$\\
$\up(3S)\to\mumu$ & $49\pm 12$  & $2.60$ & $0.71\pm 0.17$ &$10.33<\mm(\pnpn)<10.39$ & 103 & $0.43$\\
$\up(3S)\to\ee$ & $9\pm 14$  & $1.19$ & $0.29\pm 0.44$ & not used & --- & ---\\
\hline\hline
\end{tabular*}
\end{table*}

Table~\ref{tab:syst_br} shows the dominant sources of systematic uncertainties 
for the cross section measurements.
The uncertainty on the data/MC difference is estimated by varying the 
requirements on $P$, $|\de|$, $\mm(\lplm)$ and $\chi^2(\pi^0)$. 
We obtain a 4\% uncertainty on both $\up(1,2S)\pnpn$ samples.
The same value is used for $\up(3S)\pnpn$ due to the small sample size 
in this final state. The uncertainty on the signal and background 
PDFs in the fit is estimated by variation of the fit range and changing the 
parameterization to a single Gaussian for the signal and a third- and 
foth-order polynomial for the background.
The systematic uncertainties on lepton ID are estimated using 
the process $\upf\to\upns\pi^+\pi^-$, $\upns\to\lplm$. 
The tracking uncertainty is obtained from partially and fully reconstructed 
$D^*\to\pi^+D^0$, $D^0\to K_S^0\pi^+\pi^-$ decays.
The $\pi^0$ reconstruction uncertainty is estimated using 
$\tau^-\to\pi^-\pi^0\nu_\tau$.
The trigger efficiency is determined by MC to be 94-99\%, depending on 
the final state. We conservatively estimate its error as 2\%.
The uncertainty of the ISR correction factor is determined by the 
modification of the parameterization of the $\ee\to\upns\pipi$ cross section
(variation of $\upf$ mass and width within its errors, including a possible 
contribution of the non-resonant term) and variation of selection criteria.
We combine different $\upns$ decay modes assuming a 100\% correlation for 
all sources of systematic errors except lepton ID. 
The total systematic errors are $8.6\%$, $12.3\%$  and $19.2\%$ 
for $\upns\pnpn$, $n=1,2$ and 3, respectively.
We calculate the weighted average of $\seeuppp$ 
in the various $\upns$ decay channels and obtain~\footnote{}
\begin{eqnarray}
&& \svis(\ee\to\up(1S)\pnpn)=(0.77\pm 0.04\pm 0.07)\, {\rm pb}\, , \nonumber \\
&& \svis(\ee\to\up(2S)\pnpn)=(1.25\pm 0.08\pm 0.15)\, {\rm pb}\, , \nonumber \\
&& \svis(\ee\to\up(3S)\pnpn)=(0.66\pm 0.16\pm 0.13)\, {\rm pb}
\end{eqnarray}
and
\begin{eqnarray}
&& \sigma(\ee\to\up(1S)\pnpn)=(1.16\pm 0.06\pm 0.10)\, {\rm pb}\, , \nonumber \\
&& \sigma(\ee\to\up(2S)\pnpn)=(1.87\pm 0.11\pm 0.23)\, {\rm pb}\, , \nonumber \\
&& \sigma(\ee\to\up(3S)\pnpn)=(0.98\pm 0.24\pm 0.19)\, {\rm pb}\, .
\end{eqnarray}
These are approximately one half of the corresponding values of
$\sigma(\ee\to\upns\pipi)$~\cite{br_ypipi,y5s_3body},
consistent with the expectations from isospin conservation.
The Born cross section $\sborn$ can be obtained by multiplying by the 
vacuum polarization correction factor:
\begin{equation}
\sborn = \sigma |1 - \Pi|^2\, ,
\end{equation}
where $|1 - \Pi|^2 = 0.9286$~\cite{actis}.
The branching fractions listed in PDG can be obtained by
\begin{equation}
\br(\upf\to\upns\pnpn) = \frac{\svis(\ee\to\upns\pnpn)}{\sbb({\rm at}\, \upf)}\, ,
\end{equation}
where $\sbb({\rm at}\,\upf)=(0.340\pm 0.016)$~nb~\cite{sigma_y5s}.

\begin{table*}
\caption{Systematic uncertainties for the cross section measurements (in \%)}
\medskip
\label{tab:syst_br}

\begin{tabular*}{\textwidth}{@{\extracolsep{\fill}}lccccccc}\hline\hline
Source & $\up(1S)[\mumu]$ & $\up(1S)[\ee]$ & $\up(2S)[\mumu]$ & $\up(2S)[\ee]$ & $\up(2S)[\up\pi^+\pi^-]$ & $\up(3S)[\mumu]$ & $\up(3S)[\ee]$\\
\hline
Data/MC difference & $4.0$ & $4.0$ & $4.0$ & $4.0$ & $4.0$ & $4.0$ & $4.0$\\
Signal/background PDF &  $3.0$ & $3.0$ & $4.0$ & $5.0$ & $5.0$ & $15.0$ & $50$\\
$\br(\upns\to X)$ \cite{PDG} & $2.0$ & $4.6$ & $8.8$ & $8.4$ & $3.3$ & $9.6$ & $9.6$\\
Leptons ID & $1.0$ & $3.0$ & $1.0$ & $3.0$ & $2.5$ & $1.0$ & $3.0$\\
Tracking & $0.7$ & $0.7$ & $0.7$ & $0.7$ & $1.7$ & $0.7$  & $0.7$\\
$\pi^0$'s reconstruction& $5.0$ & $5.0$ & $5.0$ & $5.0$ & $5.0$ & $5.0$ & $5.0$\\
Luminosity & $1.4$ & $1.4$ & $1.4$ & $1.4$ & $1.4$ & $1.4$ & $1.4$\\
Trigger efficiency & $2.0$ & $2.0$ & $2.0$ & $2.0$ & $2.0$ & $2.0$ & $2.0$\\ 
$\disr$ & $2.0$ & $2.0$ & $2.0$ & $2.0$ & $2.0$ & $2.0$ & $2.0$ \\ \hline
Sum for $\svis$ & $7.8$ & $9.3$ & $11.9$ & $12.3$ & $9.6$ & $19.1$ & $52$\\
\hline
Sum for $\sigma$& $8.1$ & $9.5$ & $12.1$ & $12.5$ & $9.8$ & $19.2$ & $52$\\
\hline\hline
\end{tabular*}
\end{table*}

\section{Dalitz Analysis}
\begin{figure*}
  \includegraphics[width=0.32\textwidth] {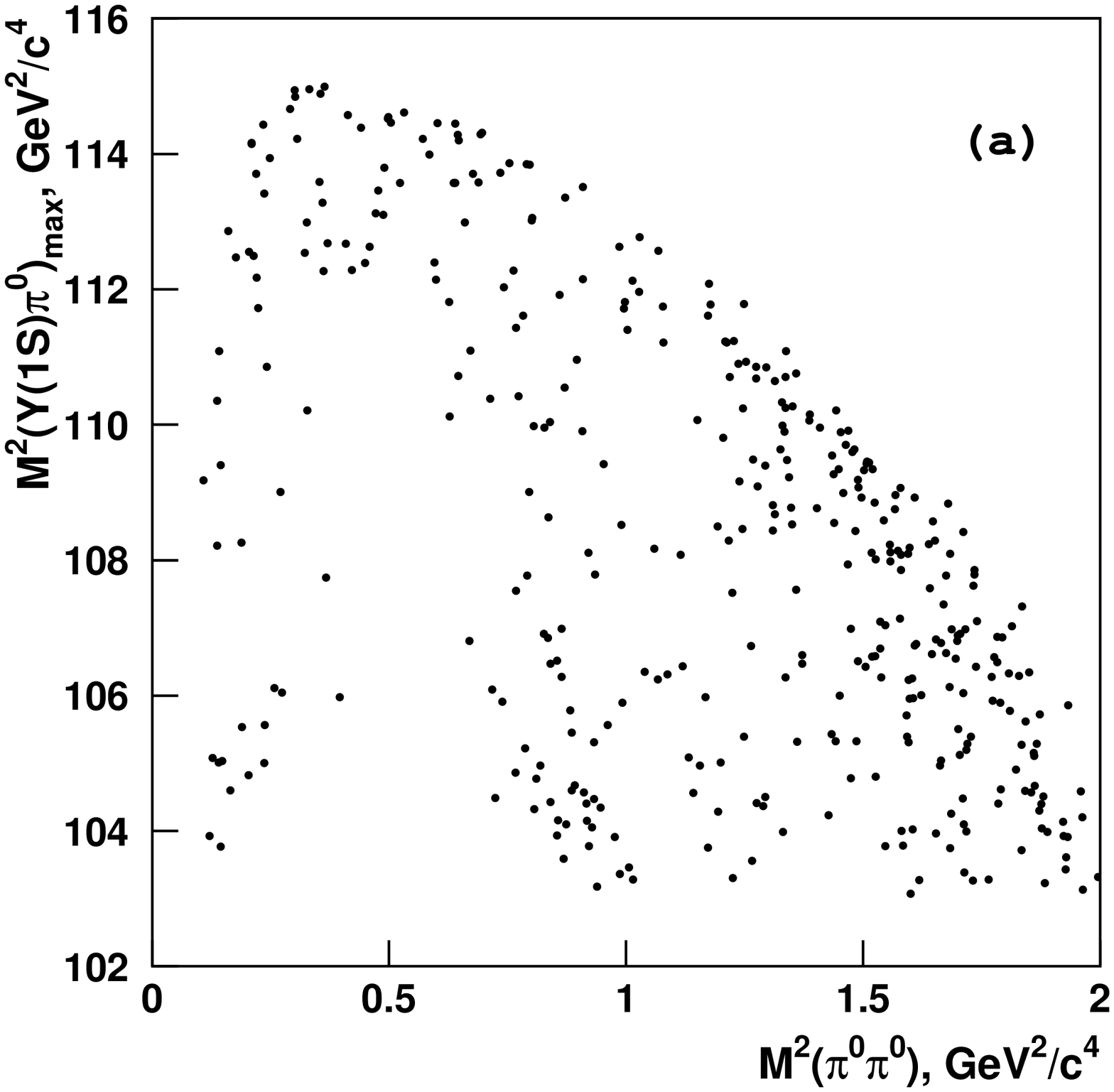}
  \includegraphics[width=0.32\textwidth] {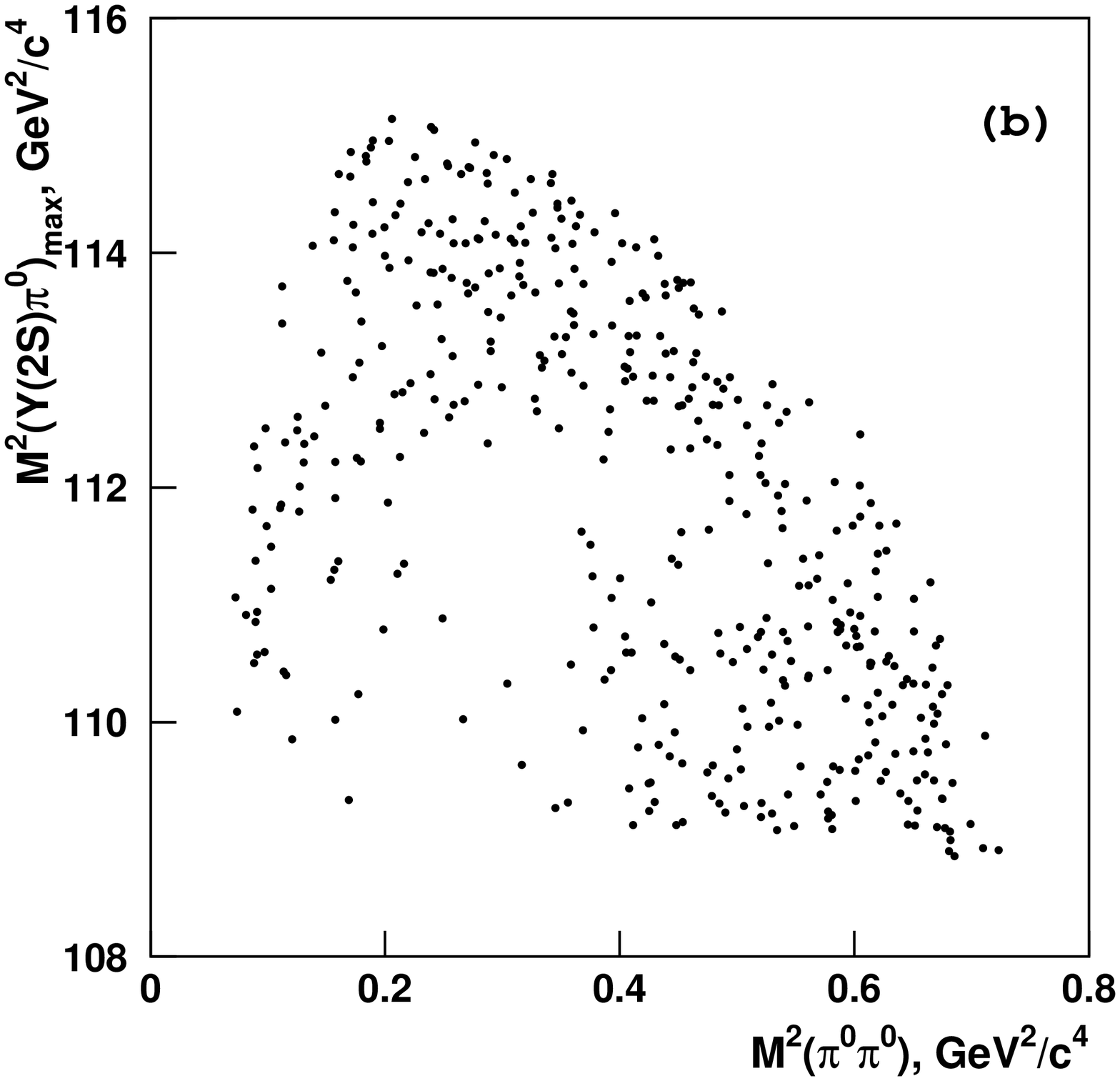}
  \includegraphics[width=0.32\textwidth] {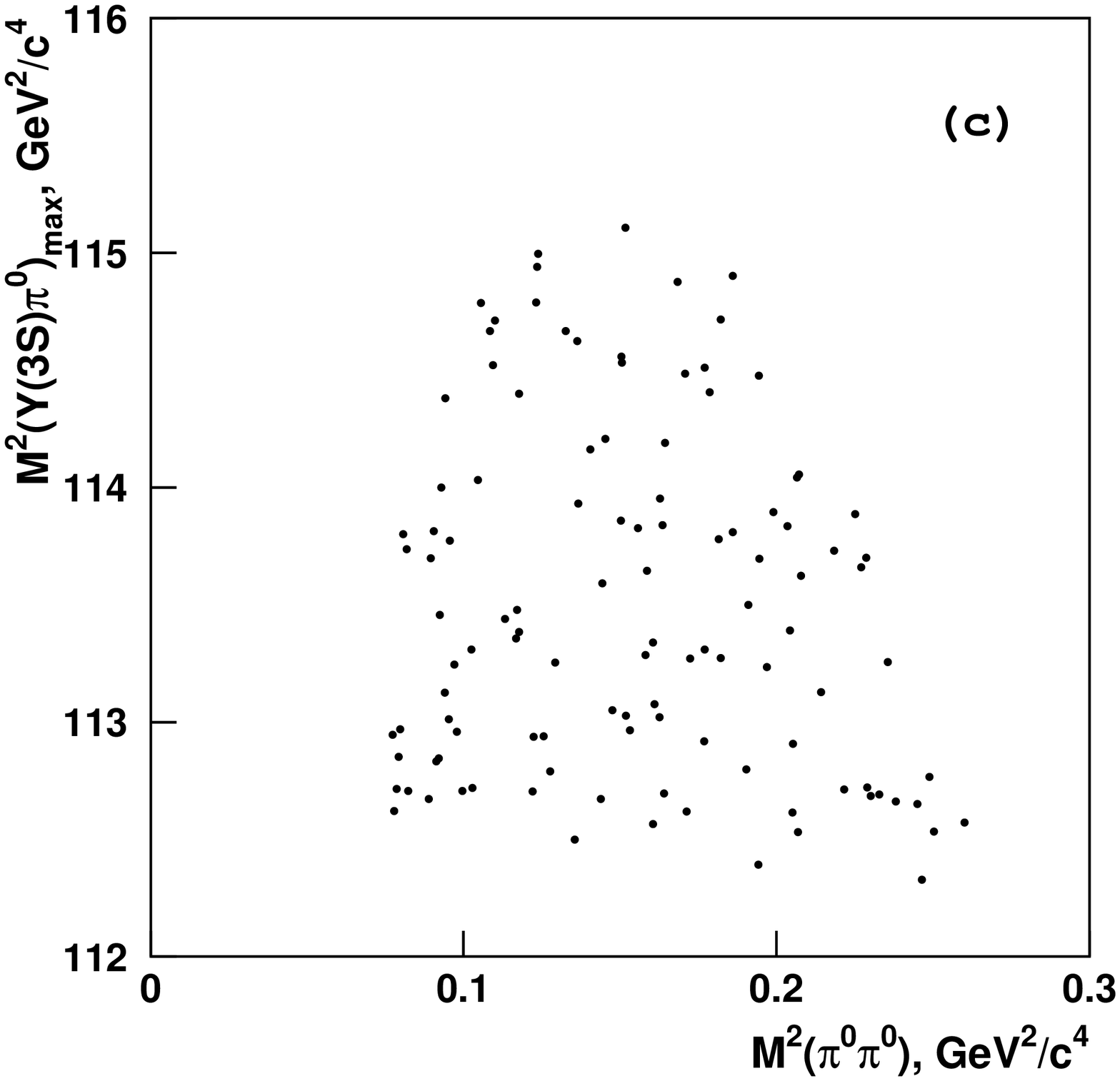}
  \caption{Dalitz plots for selected (a) $\up(1S)\pnpn$, (b) $\up(2S)\pnpn$ and (c) $\up(3S)\pnpn$
    candidates.}
  \label{fig:dalitz}
\end{figure*}

Figure~\ref{fig:dalitz} shows the Dalitz distributions for the selected
$\upf\to\upns\pnpn$ candidates in the signal regions given in Table~\ref{tab:yield}.
A mass-constrained fit is performed for the $\upns$ candidates.
Samples of background events are selected in the $\mm(\pnpn)$ sidebands 
for $\upns\to \lplm$ and in the $M(\up(1S)\pipi)$ sidebands for $\up(2S)\to \up(1S)\pipi$. 
Then we refit candidates to the nominal mass of the corresponding $\upns$ 
state to match the phase space boundaries. 
We use the following sideband regions:
$[9.20:9.35]\, \gevcc$ and $[9.60:9.75]\, \gevcc$ for $\up(1S)[\lplm]\pnpn$;
$[9.80:9.95]\, \gevcc$ and $[10.15:10.30]\, \gevcc$ for $\up(2S)[\lplm]\pnpn$;
$[9.90:9.95]\, \gevcc$ and $[10.10:10.20]\, \gevcc$ for $\up(2S)[\up\pipi]\pnpn$;
$[10.15:10.30]\, \gevcc$ and $[10.45:10.50]\, \gevcc$ for $\up(3S)[\lplm]\pnpn$.
We parameterize the background PDF by the following function:
\begin{equation}
B(s_1,s_2) = 1+p_1 e^{-q_1 s_3} + p_2 e^{-q_2 (s_{\rm min}-c_2)},
\end{equation}
where $p_1$, $p_2$, $q_1$ and $q_2$ are the fit parameters,
$s_3=M^2(\pnpn)$, $s_{\rm min}={\rm min}(s_1, s_2)$ and 
$s_{1,2}=M^2(\upns\pi^0_{1,2})$,
$c_2$ is defined as $(m_{\upns}+m_{\pi^0})^2$.

Variation of the reconstruction efficiency over the Dalitz plot is 
determined using a large sample of MC with a uniform phase space 
distribution.
We use the following function to parameterize the efficiency:
\begin{equation}
\epsilon=1+\alpha\{1-e^{-(s_3-c_0)/b_0}\}\{1-e^{-(c_1-s_{\rm max})/b_1}\},
\end{equation}
where $\alpha$, $b_0$ and $b_1$ are fit parameters,
$s_{\rm max}={\rm max}(s_1, s_2)$,
$c_0$ and $c_1$ are defined as 
$c_0=4m^2_{\pi^0}$ and $c_1 = (m_{\upf}-m_{\pi^0})^2$.

The amplitude analysis of the three-body $\upf\to\upns\pnpn$ decays 
closely follows Ref.~\cite{zb_paper}.
We describe the three-body signal amplitude with a sum of quasi-two-body 
contributions:
\begin{equation}
{\mathcal M}(s_1,s_2)=A_{Z1}+A_{Z2}+A_{f_0}+A_{f_2}+\anr\ ,
\end{equation}
where $A_{Z1}$ and $A_{Z2}$ are the amplitudes for contributions from the 
$\zbnf$ and $\zbns$, respectively. 
The amplitudes $A_{f_0}$, $A_{f_2}$ and $\anr$ account 
for the contributions from the $\pnpn$ system in an $f_0(980)$, $f_2(1275)$ 
and a non-resonant state, respectively.
We assume that the dominant contributions to $A_{Zk}$ are from amplitudes 
that preserve the orientation of the spin of the heavy quarkonium state 
and, thus, both pions in the cascade decay $\upf\to\zbn\pi^0\to\upns\pnpn$ 
are emitted in an $S$-wave with respect to the heavy quarkonium system. 
As demonstrated in Ref.~\cite{zb_helicity}, angular analysis supports 
this assumption. Consequently, we parameterize both amplitudes with an 
$S$-wave Breit-Wigner function, neglecting the possible $s$ dependence 
of the resonance width:
\begin{equation}
\label{eq:bw}
{\rm BW}(s,M,\Gamma)=\frac{\sqrt{M\Gamma}}{M^2-s-iM\Gamma}\,.
\end{equation}
Both amplitudes are symmetrized with respect to $\pi^0$ interchange:
\begin{equation}
A_{Zk}(k=1,2)=a_ke^{i\delta_k}(BW(s_1,m_k,\Gamma_k)+BW(s_2,m_k,\Gamma_k))\,.
\end{equation}
The masses and widths are fixed to the 
values obtained in the $\upns\pipi$ and $h_b(mP)\pipi$ analyses: 
$M(Z_1)=10607.2\,\mevcc$, 
$\Gamma(Z_1)=18.4\,\mev$, $M(Z_2)=10652.2\,\mevcc$ and
$\Gamma(Z_2)=11.5\,\mev$~\cite{zb_paper}.
We use a Flatt\'{e} function~\cite{flatte} for the $f_0(980)$ and a 
Breit-Wigner function for the $f_2(1275)$.
Coupling constants of the $f_0(980)$ are fixed at the values from 
the $B^+\to K^+\pipi$ analysis: $M=950\,\mevcc$, $g_{\pi\pi}=0.23$ and
$g_{KK}=0.73$~\cite{kpipi}. 
The mass and width of the $f_2(1275)$ resonance are fixed to the world average 
values~\cite{PDG}.
Following suggestions in Ref.~\cite{voloshin}, the non-resonant
amplitude $\anr$ is parameterized as 
\begin{equation}
\anr=\anr_1 e^{i\phinr_1} + \anr_2 e^{i\phinr_2} s_3\ ,
\end{equation}
where $\anr_1$, $\anr_2$, $\phinr_1$ and $\phinr_2$ 
are free parameters in the fit.
As there is only sensitivity to the relative amplitudes and
phases between decay modes, we fix $\anr_1=10.0$ and $\phinr_1=0.0$.
Since the phase space of the decay $\upf\to\up(3S)\pnpn$ is very limited,
contributions from $f_0$ and $f_2$ are not included in the fit.

We perform an unbinned maximum likelihood fit.
The likelihood function is defined as
\begin{equation}
{\mathcal L}=\prod\epsilon(s_1,s_2)\,\bigl(\fsig S(s_1,s_2)+(1-\fsig) B(s_1,s_2)\bigr)\ ,
\end{equation}
where the product runs over all signal candidates.
$S(s_1,s_2)$ is $|{\mathcal M}(s_1,s_2)|^2$ convoluted with the detector 
resolution ($6.0\,\mevcc$ for $M(\upns\pi^0)$); $\epsilon(s_1,s_2)$ describes 
the variation of the reconstruction efficiency over the Dalitz plot.
The fraction $\fsig$ is the fraction of signal events in the data sample
determined separately for each $\upns$ decay mode (see Table~\ref{tab:yield}).
The function $B(s_1,s_2)$ describes the distribution of background events 
over the phase space. Both products $S(s_1,s_2)\epsilon(s_1,s_2)$ and 
$B(s_1,s_2)\epsilon(s_1,s_2)$ are normalized to unity.

\begin{figure*}
  \includegraphics[width=0.32\textwidth] {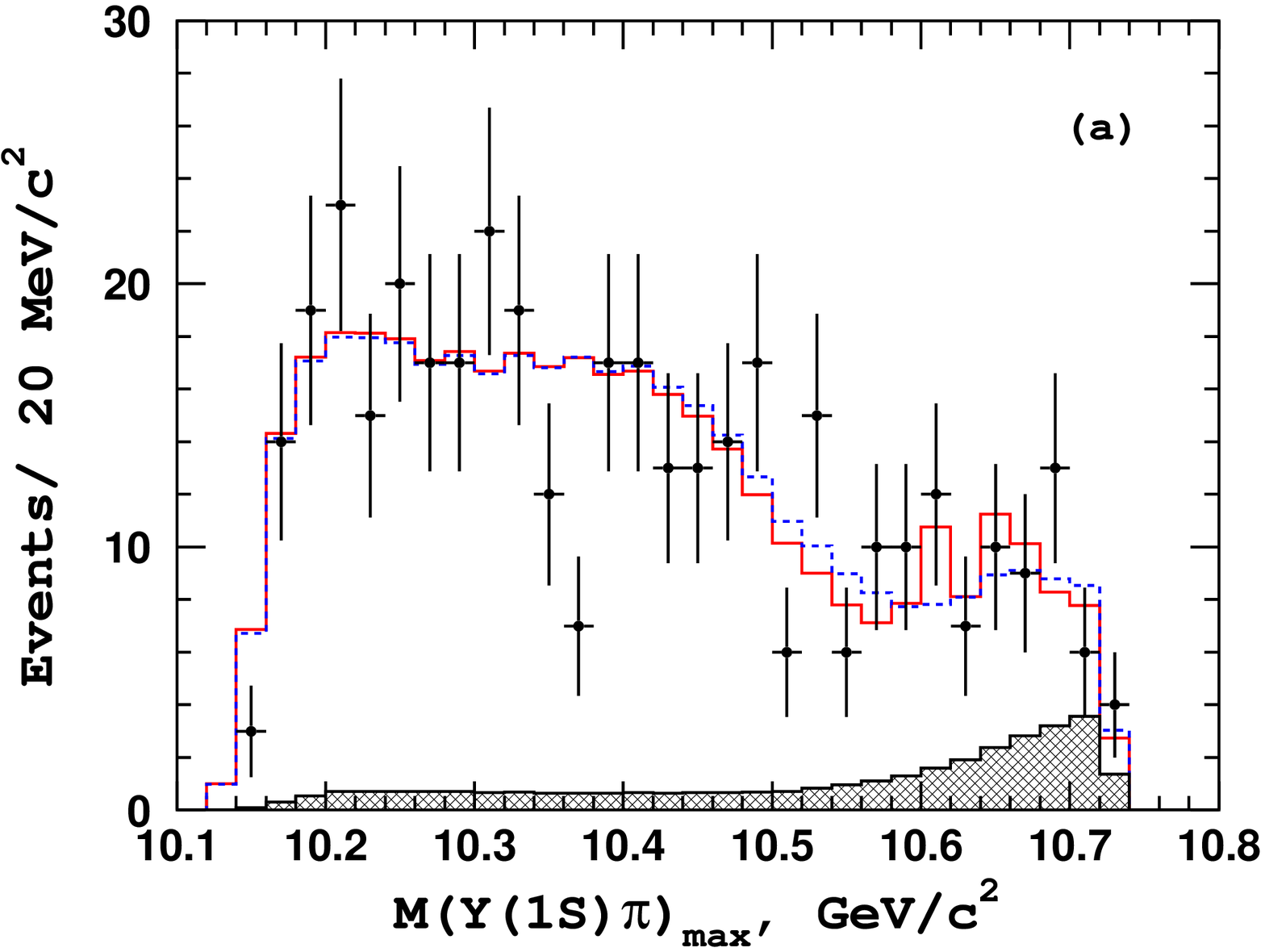}
  \includegraphics[width=0.32\textwidth] {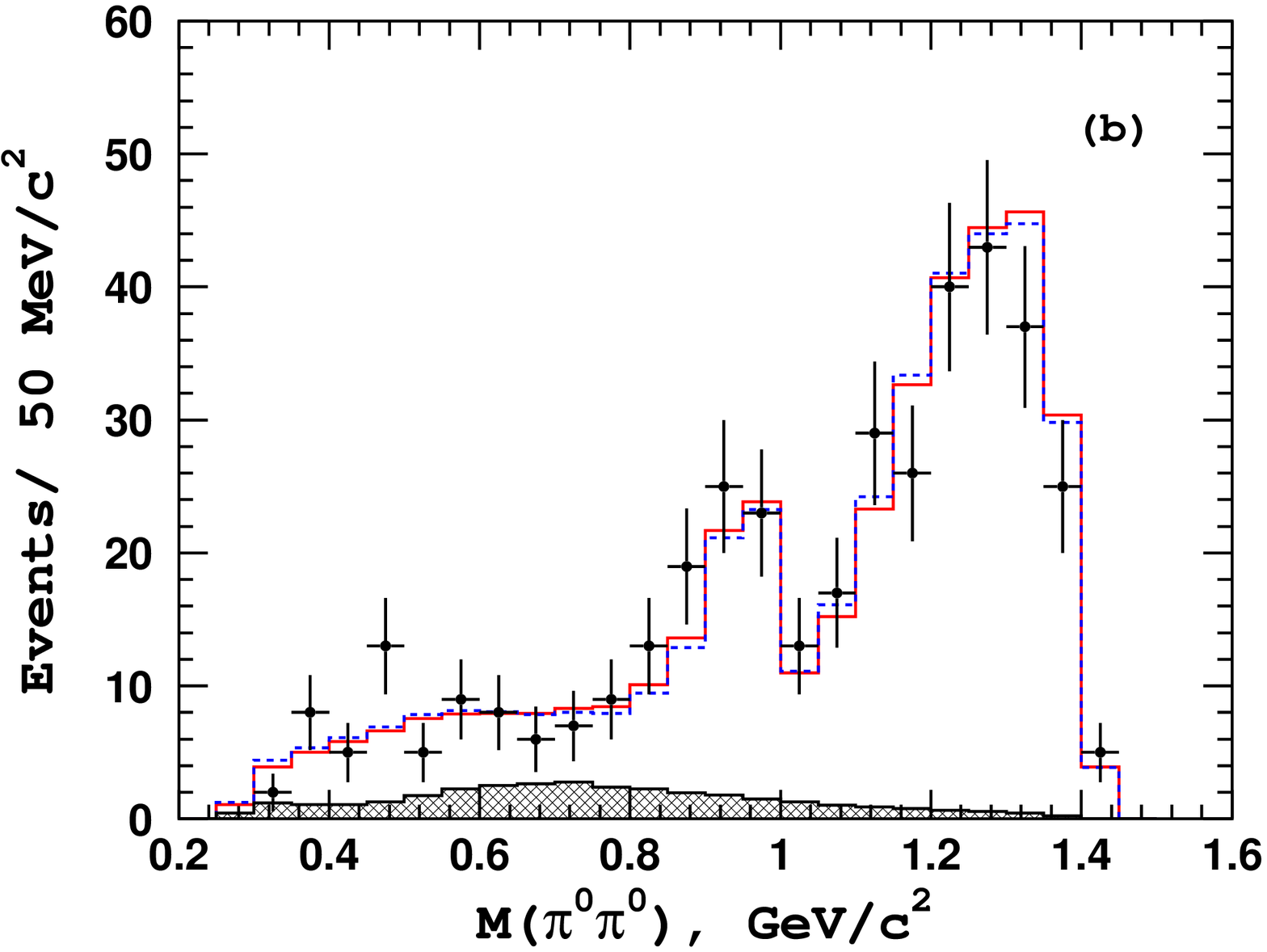}
  \includegraphics[width=0.32\textwidth] {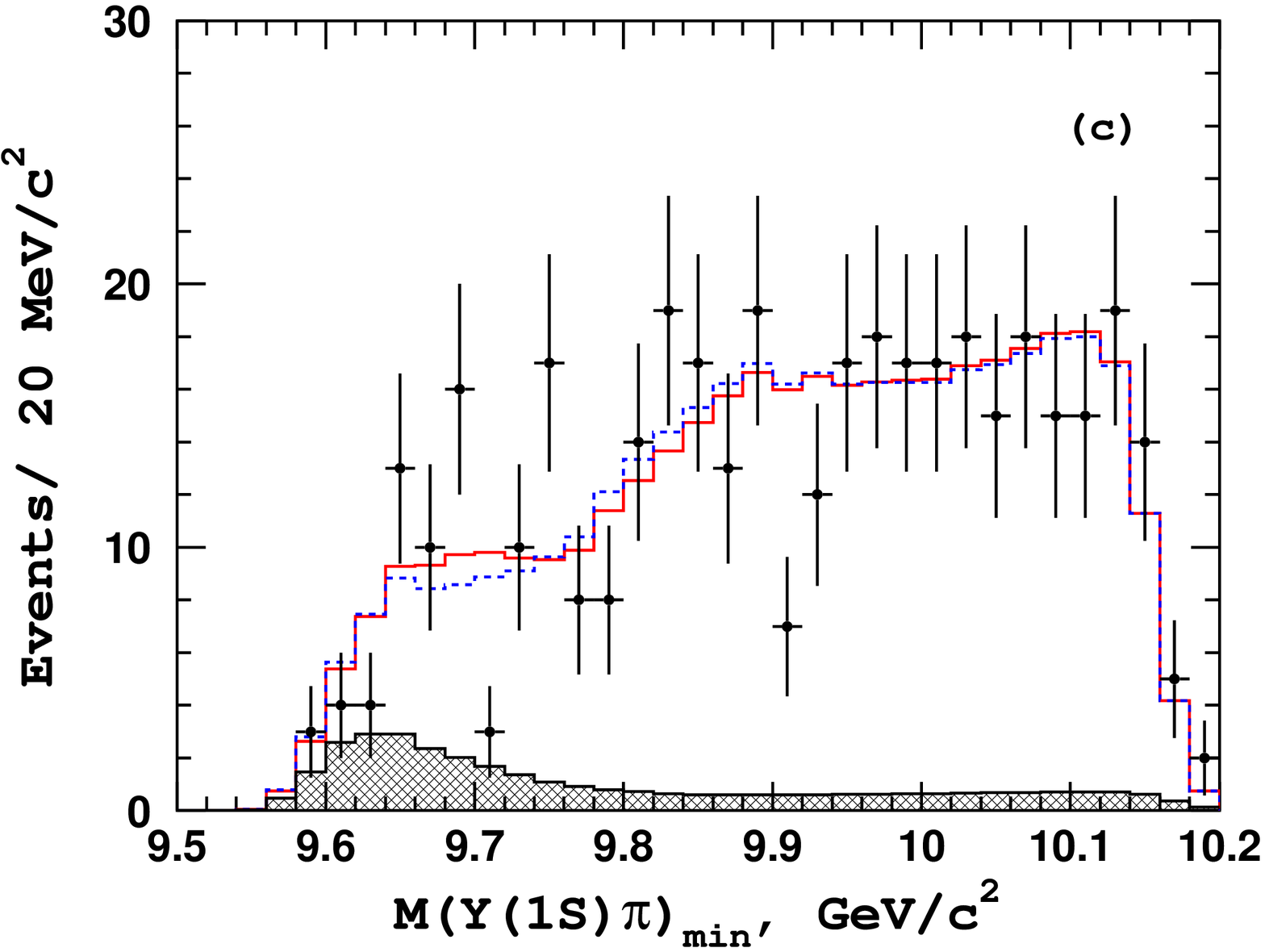}
  \caption{Comparison of the fit results (open histograms) with experimental 
    data (points with error bars) for $\up(1S)\pnpn$ events in the signal 
    region. 
    Solid red and dashed blue open histograms show the fit with and without 
    $\zbn$'s, respectively. Hatched histograms show the background components.}
  \label{fig:y1s_fitres}
\end{figure*}

\begin{figure*}
  \includegraphics[width=0.32\textwidth] {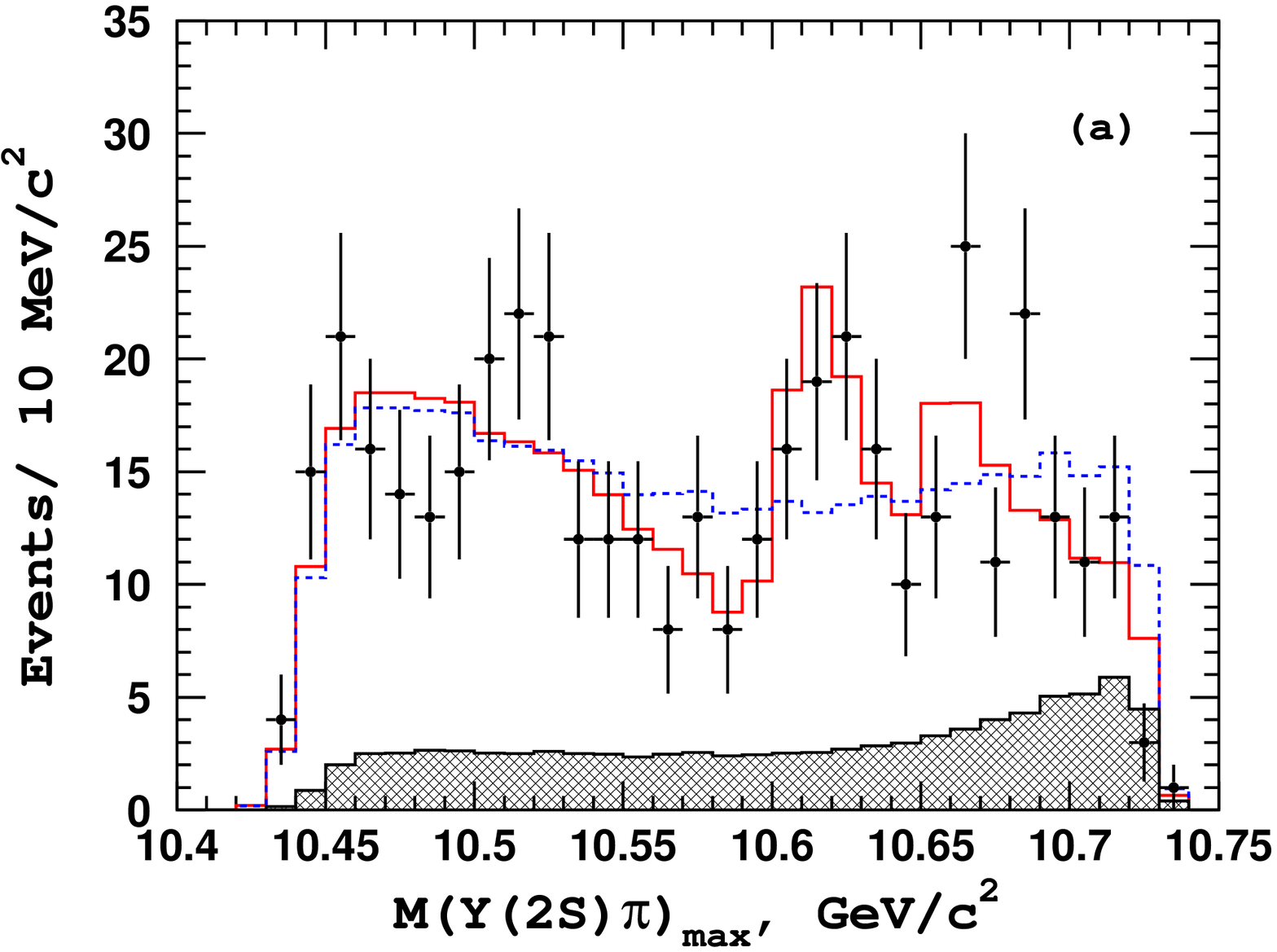}
  \includegraphics[width=0.32\textwidth] {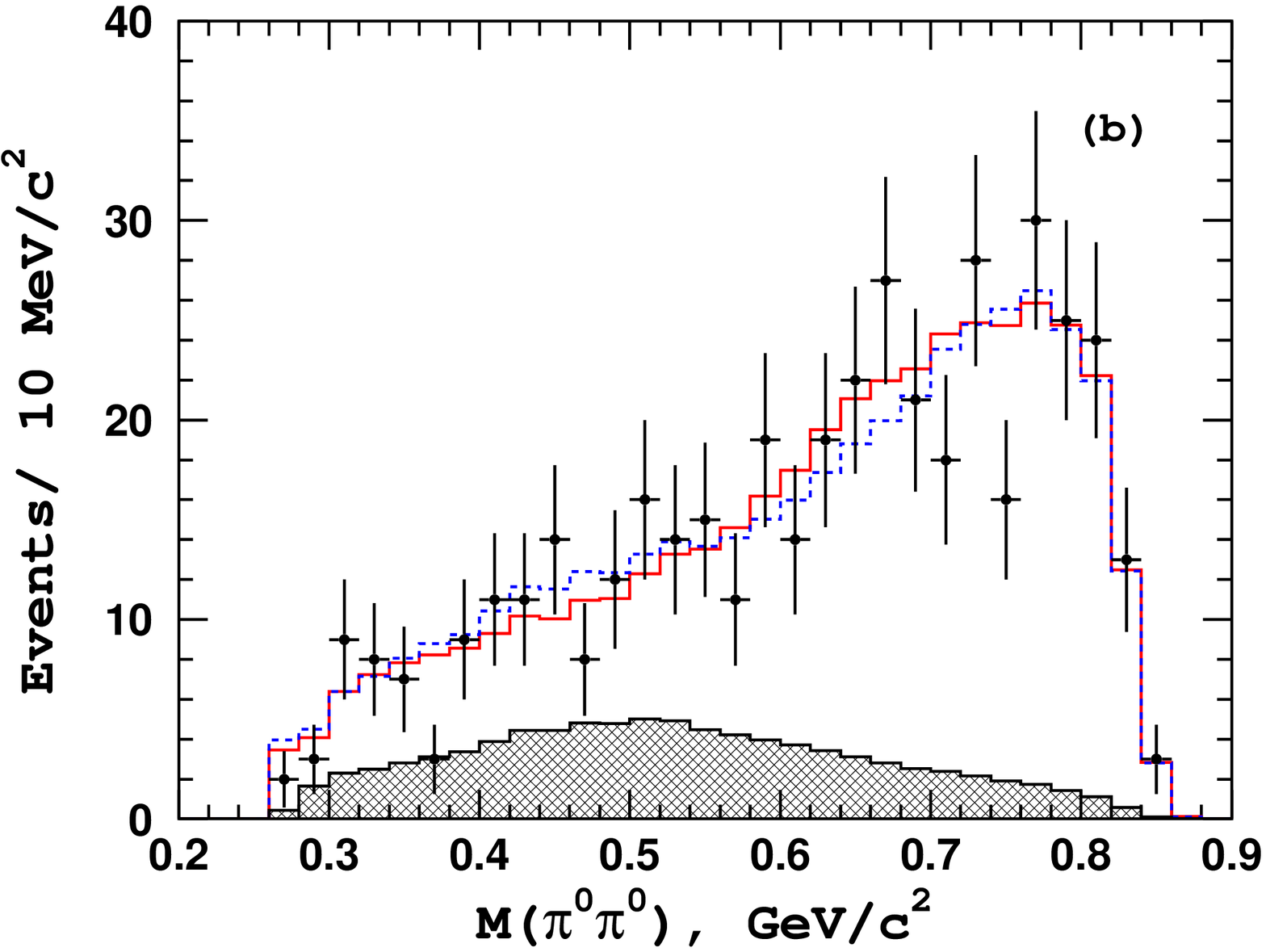}
  \includegraphics[width=0.32\textwidth] {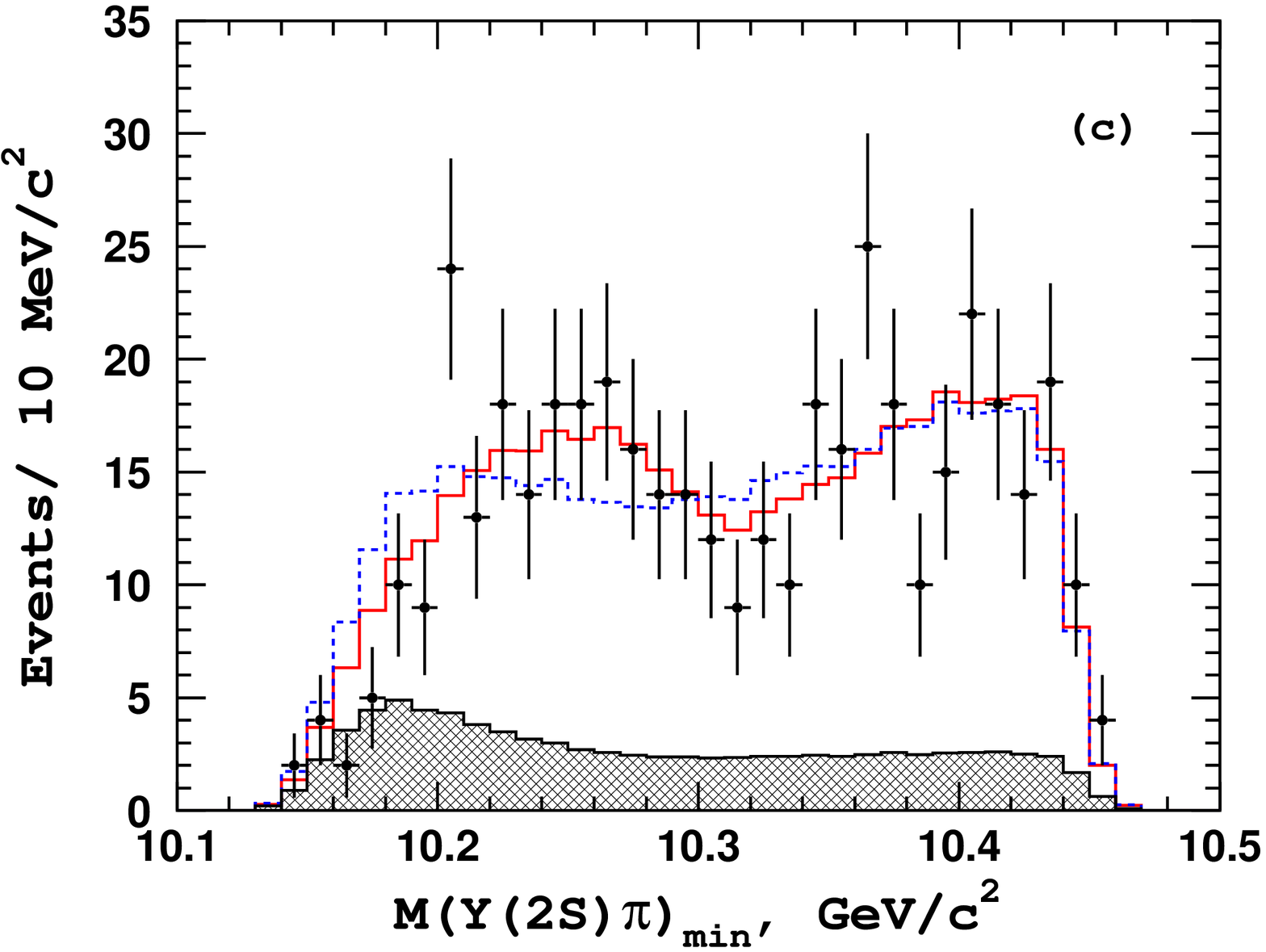}
  \caption{Comparison of the fit results (open histograms) with experimental 
    data (points with error bars) for $\up(2S)\pnpn$ events in the signal 
    region. The legends are the same as in Fig.~\ref{fig:y1s_fitres}.
    Only solution A is shown. Both solutions give indistinguishable plots.}
  \label{fig:y2s_fitres}
\end{figure*}

\begin{figure*}
  \includegraphics[width=0.32\textwidth] {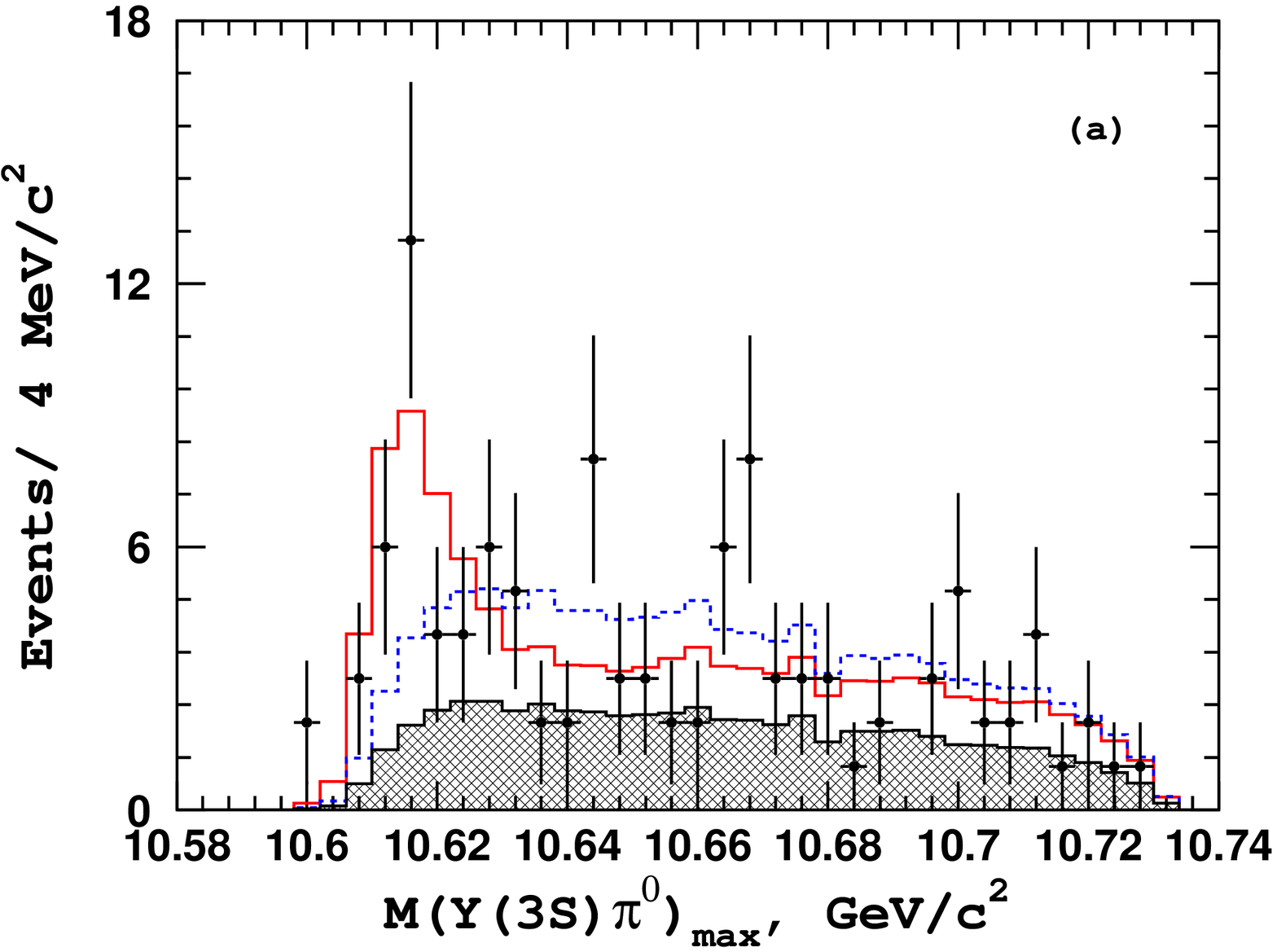}
  \includegraphics[width=0.32\textwidth] {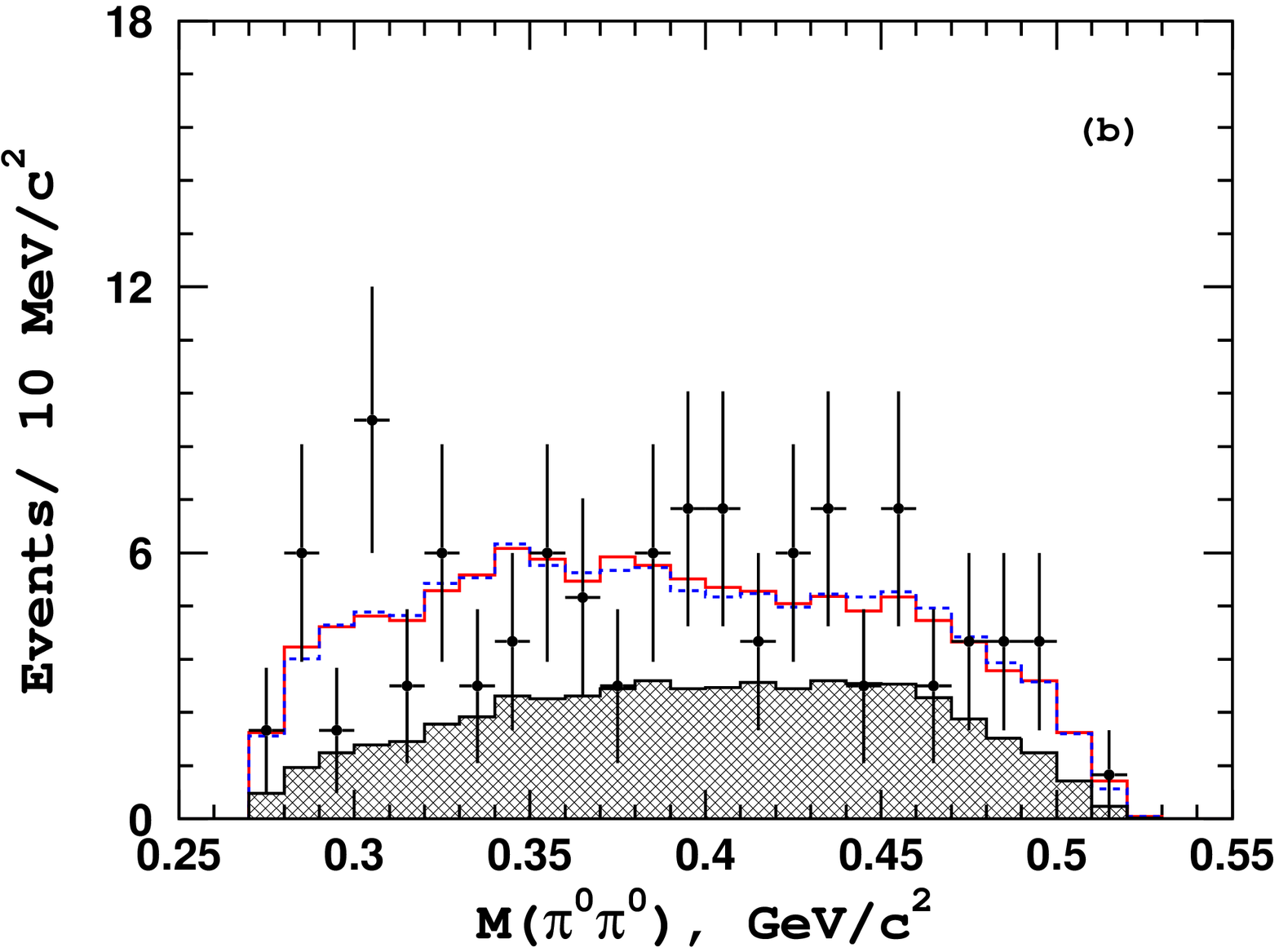}
  \includegraphics[width=0.32\textwidth] {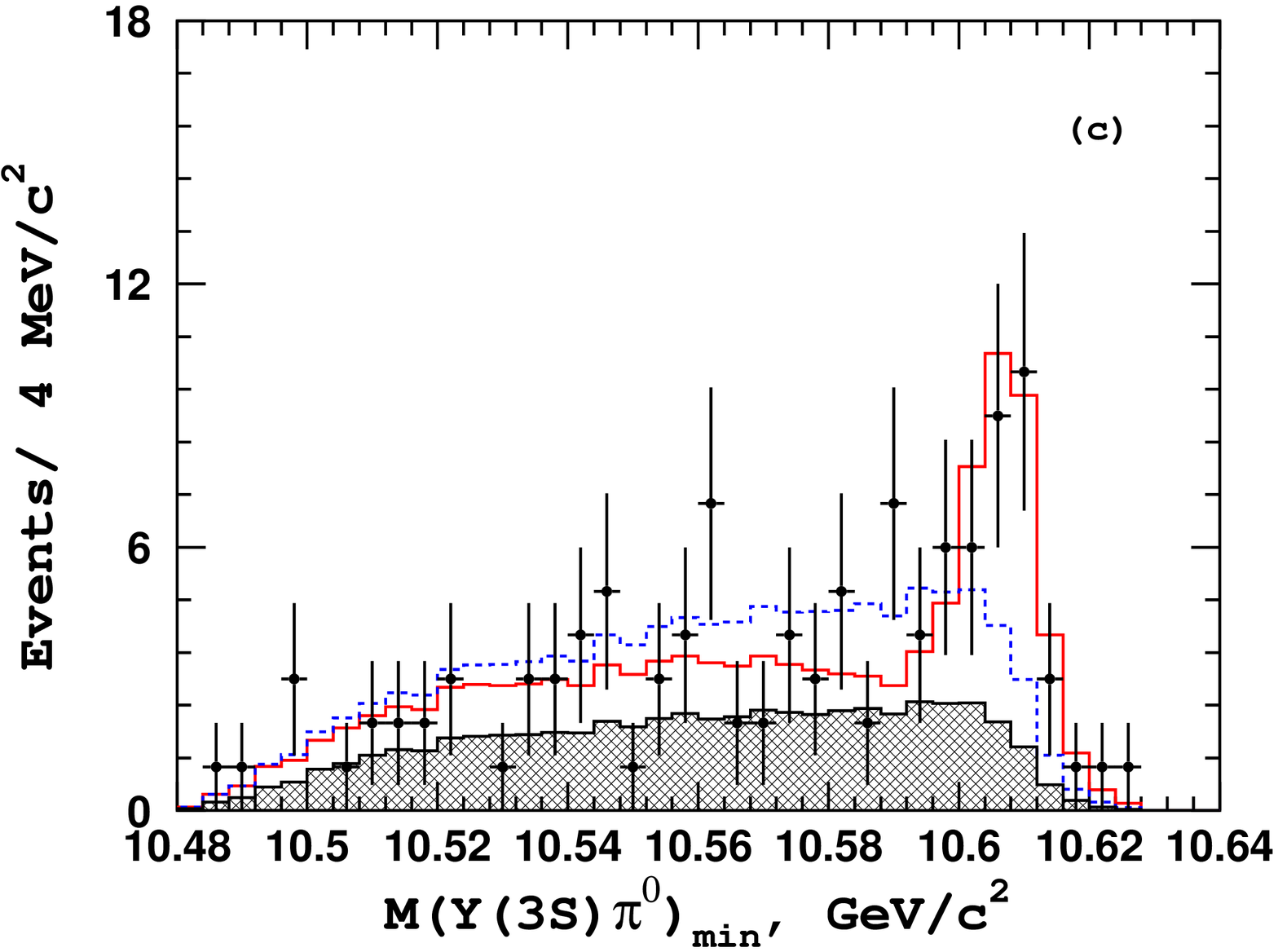}
  \caption{Comparison of the fit results (open histograms) with experimental 
    data (points with error bars) for $\up(3S)\pnpn$ events in the signal 
    region. The legends are the same as in Fig.~\ref{fig:y1s_fitres}.}
  \label{fig:y3s_fitres}
\end{figure*}

To ensure that the fit converges to the global minimum, we perform $10^3$ 
fits with randomly assigned initial values for amplitudes and phases.
We find two solutions for the $\up(2S)\pnpn$ sample with similar values of
$-\tl$ (see Table \ref{tab:mulsol}). Solution A has better consistency
with the Dalitz plot fit result for the $\upf\to\up(2S)\pipi$ 
decay~\cite{y5s_3body}. 
We find single solutions for the $\up(1,3S)\pnpn$ samples.
\begin{table*}
\caption{Two solutions found in the Dalitz plot fit of $\up(2S)\pnpn$ 
events. The phases are in degrees. The non-resonant amplitude $\anr_1$ and its 
phase are fixed to $10.0$ and $0.0$, respectively.}
\medskip
\label{tab:mulsol}
\begin{tabular*}{\textwidth}{@{\extracolsep{\fill}}lcccccc}\hline\hline
 & w/o $\zbn$ & with $Z_1^0$ & with $\zbn$'s & w/o $\zbn$ & with $Z_1^0$ & with $\zbn$'s\\
Solutions & A & A & A & B & B & B\\ 
\hline
$A(Z_1^0)$   & $0.0$ \fix   &$0.46^{+0.15}_{-0.11}$&$0.58_{-0.14}^{+0.21}$&  $0.0$ \fix    &$1.35^{+0.64}_{-0.33}$&$1.42\pm 0.48$\\
$\phi(Z_1^0)$& ---          & $243\pm 14$         & $247\pm 14$         &  ---           & $88\pm 18$           & $91\pm 21$ \\
$A(Z_2^0)$   & $0.0$ \fix   & $0.0$ \fix           &$0.37_{-0.16}^{+0.20}$&  $0.0$ \fix    &  $0.0$ \fix          & $0.66\pm 0.40$\\
$\phi(Z_2^0)$& ---          & ---                  & $235\pm 27$         &  ---           &  ---                 & $124\pm 37$ \\
\hline                                                                    
$A(f_2)$     & $28.2\pm 7.0$& $23.9\pm 7.3$        & $18.2\pm 7.3$        & $41.8\pm 9.0$  & $48.7\pm 15.4$       & $43.3\pm 15.6$\\
$\phi(f_2)$  & $28\pm 10$   & $28\pm 13$           & $36\pm 21$           & $359\pm 14$     & $10\pm 16$           & $132\pm 19$\\
$A(f_0)$     & $8.2\pm 2.1$ & $10.5\pm 1.9$        & $11.5\pm 1.9$        & $13.3\pm 3.6$  & $13.4\pm 4.2$        & $12.6\pm 4.9$\\
$\phi(f_0)$  & $210\pm  8$  & $213\pm 7$           & $211\pm 6$           & $131\pm 11$    & $134\pm 15$          & $132\pm 19$    \\
$\anr_2$     & $24.6\pm 4.2$& $31.8\pm 4.3$        & $34.7\pm 4.9$        & $44.2\pm 10.1$ & $50.4\pm 12.2$       & $50.8\pm 13.7$ \\
$\phinr_2$   & $93\pm 15$   & $85\pm 13$           & $80\pm 12$           & $290\pm 16$    & $291\pm 22$          & $288\pm 25$ \\
\hline                                                                    
$-\tl$       & $-154.5$     & $-186.6$             & $-193.1$             & $-155.4$       & $-186.3$             & $-191.2$  \\
\hline\hline
\end{tabular*}
\end{table*}
Table~\ref{tab:main_fit} shows the values and errors of amplitudes and 
phases obtained from the fit to the $\up(1S)\pnpn$ and $\up(3S)\pnpn$ Dalitz plots.
\begin{table*}
\caption{Results of the Dalitz plot fit of $\up(1,3S)\pnpn$ events.
The phases are in degrees. The non-resonant amplitude $\anr_1$ and its phase 
are fixed to $10.0$ and $0.0$, respectively.}
\medskip
\label{tab:main_fit}
\begin{tabular*}{\textwidth}{@{\extracolsep{\fill}}lccccc}\hline\hline
     & $\up(1S)\pnpn$ & $\up(1S)\pnpn$ & $\up(3S)\pnpn$ & $\up(3S)\pnpn$& $\up(3S)\pnpn$\\
Model & with $\zbn$'s & w/o $\zbn$'s   & with $\zbn$'s  &with $Z_1^0$ only & w/o $\zbn$'s\\ 
\hline
$A(Z_1^0)$ &$0.50_{-0.30}^{+0.34}$& $0.0$ \fix&$1.07_{-0.33}^{+1.45}$&$1.09_{-0.31}^{+0.75}$& $0.0$\fix \\
$\phi(Z_1^0)$& $324\pm 50$   & ---        & $158\pm 25$      & $149\pm 24$        & ---       \\
$A(Z_2^0)$& $0.60_{-0.47}^{+0.51}$& $0.0$ \fix& $0.32_{-0.32}^{+1.18}$& $0.0$ \fix         & $0.0$ \fix\\
$\phi(Z_2^0)$& $301\pm 60$  & ---         & $252\pm 81$      & ---                & ---       \\
\hline
$A(f_2)$   & $15.7\pm 2.0$& $14.6\pm 1.6$& $0.0$ \fix & $0.0$ \fix & $0.0$ \fix\\
$\phi(f_2)$& $60\pm 11$   & $51\pm 9$    & ---        & ---        & ---        \\
$A(f_0)$   &$1.07\pm 0.15$&$0.97\pm 0.12$& $0.0$ \fix & $0.0$ \fix & $0.0$ \fix\\
$\phi(f_0)$& $168\pm 11$  & $163\pm 10$  & ---        & ---        & ---        \\
$\anr_2$   & $15.2\pm 1.2$& $13.9\pm 0.7$& $50.5\pm 14.1$ & $44.8\pm 12.5$ & $48.0\pm 12.7$\\
$\phinr_2$ & $162\pm 4$   & $161\pm 4$   & $155\pm 15$    & $153\pm 14$   & $151\pm 15$\\
\hline
$-\tl$     & $-316.7$     & $-312.4$     & $-31.3$        & $-30.7$       & $-5.3$    \\
\hline\hline
\end{tabular*}
\end{table*}
Projections of the fits are shown in Figs.~\ref{fig:y1s_fitres}-\ref{fig:y3s_fitres}.
These projections are very similar to the corresponding distributions 
in $\upns\pipi$~\cite{zb_paper}.
The $\zbn$ signal is most clearly observed in $M(\up(2S)\pi^0)_{\rm max}$,
$M(\up(3S)\pi^0)_{\rm max}$ and $M(\up(3S)\pi^0)_{\rm min}$.

The $\zbn$ significance is calculated from a large number of 
pseudo-experiments, each with the same statistics as in data.
MC samples are generated using models without the $\zbn$ contribution.
We fit them with and without the $\zbnf$ contribution and
examine the $\dtl$ distributions.
We find $5.3\sigma$ for the $\zbnf$ statistical significance in both 
solutions for $\up(2S)\pnpn$. In addition, the $\zbnf$ statistical 
significance is $4.7\sigma$ in the fit to the $\up(3S)\pnpn$ sample.
The $\zbnf$ signal is not significant in the fit to the $\up(1S)\pnpn$ events
due to the smaller relative branching fraction. The signal for the  
$\zbns$ is not significant in any of the $\up(1,2,3S)\pnpn$ datasets. 

We calculate the relative fit-fraction of each resonance as the ratio 
$f_{R}=\frac{\int_{\mathrm Dalitz}{|M^2_{\mathrm A_R}|}}{\int_{\mathrm Dalitz}{|M^2_{\mathrm all}|}}$
from the central values of the fit given in Tables~\ref{tab:mulsol} 
and \ref{tab:main_fit}.
Errors and 90\% CL upper limits for non-significant fractions 
are obtained using pseudo-experiments.
Results are summarized in Table~\ref{tab:zbfrac}.
The sum of individual contributions is not equal to 100\% due to 
interference between amplitudes.
Reasonable agreement is observed with the corresponding fit-fractions in the 
$\upns\pipi$ analysis~\cite{y5s_3body}.
Table~\ref{tab:zb_upf} shows the product of cross sections and branching fractions
$\sigma(\ee\to\zbn\pi^0)\cdot\br(\zbn\to\upns\pi^0)$.

\begin{table*}
\caption{Summary of results for the fit-fractions of individual channels 
in the $\upns\pnpn$ final state.}
\medskip
\label{tab:zbfrac}
\begin{tabular*}{\textwidth}{@{\extracolsep{\fill}}lcccc}\hline\hline
Fraction, \% & $\up(1S)$ & $\up(2S)$ solution A & $\up(2S)$ solution B & $\up(3S)$\\
\hline
$\zbnf$ & $0.9^{+2.2+0.5}_{-0.9-0.3} (<4.6)$ & $13.5^{+6.8+3.2}_{-2.7-4.4}$ & $25.4^{+6.2+4.2}_{-5.9-11}$ & $84^{+17+14}_{-23-11}$\\
$\zbns$ & $0.6^{+2.5+0.5}_{-0.6-0.3} (<4.8)$ & $2.7^{+3.0+1.5}_{-1.4-1.2} (<8.0)$& $2.7^{+5.8+1.2}_{-1.6-1.2}(<12.4)$& $4.3^{+2.4+3.5}_{-2.2-1.9} (<10.9)$ \\
$f_2(1275)$ & $26.3\pm 4.2^{+7.8}_{-4.5}$     & $3.9^{+3.4+3.8}_{-2.0-2.1}$ & $8.7^{+4.6+3.9}_{-2.0-4.5}$  & ---  \\
Total S-wave  & $72.4\pm 4.7^{+5.6}_{-3.4}$ & $95.5^{+5.2+6.0}_{-6.2-6.5}$ & $110^{+7+6}_{-9-18}$ & $65^{+12+18}_{-15-17}$ \\
\hline
Sum & $100^{+9}_{-6}\pm 1$ & $116^{+10}_{-4}\pm 3$ & $145^{+12}_{-10}\pm 9$ & $153^{+38}_{-22}\pm 15$\\
\hline\hline
\end{tabular*}
\end{table*}

\begin{table*}
\caption{Product of the $\sigma(\ee\to\zbn\pi^0)\cdot\br(\zbn\to\upns\pi^0)$.}
\medskip
\label{tab:zb_upf}
\begin{tabular*}{\textwidth}{@{\extracolsep{\fill}}lcccc}\hline\hline
$\sigma\cdot \br$, fb & $\up(1S)$ & $\up(2S)$ solution A & $\up(2S)$ solution B & $\up(3S)$\\
\hline
$\zbnf$ & $10^{+26+6}_{-10-4} (<59)$ & $252^{+127+67}_{-52-88}$ & $475^{+119+98}_{-114-214}$ & $823^{+317+256}_{-367-234}$\\
$\zbns$ & $7^{+29+6}_{-7-4} (<62)$ & $50^{+56+28}_{-26-22} (<168)$& $50^{+108+22}_{-30-22}(<260)$& $42^{+26+35}_{-24-20} (<141)$ \\
\hline\hline
\end{tabular*}
\end{table*}

We perform a simultaneous fit of the $\up(2S)\pnpn$ and $\up(3S)\pnpn$ data 
samples. No constraints between samples are imposed on signal model 
parameters and the background description. 
The combined significance of the $\zbnf$ signal is $6.8\sigma$. 
Results for the simultaneous 
fit are exactly the same as in separate fits of $\up(2,3S)\pnpn$ samples, 
as expected. We also perform a simultaneous fit with the $\zbnf$ 
mass as a free parameter and find $m(\zbnf)=(10609\pm 4\pm 4)\,\mevcc$;
this is consistent with the mass of the $\zbc(10610)$.

\section{Systematic Uncertainties in the Dalitz Analysis}
Table \ref{tab:fracsyst} shows the main sources of
systematic uncertainties for the measurement of fractions obtained from a 
fit of individual channels.
The model uncertainty originates mainly from the parameterization 
of the non-resonant amplitude. 
Four additional models are used: 
with an additional $f_0(500)$ resonance, parameterized by a Breit-Wigner 
function with $M=600\,\mevcc$ and $\Gamma=400$~MeV$/c$;
a model with $\anr=a e^{i\phi_a} + b e^{i\phi_b} \sqrt{s(\pnpn)}$;
a model without the $f_0(980)$ contribution;
and a model without the $\anr_2$ contribution.
Another source of systematic uncertainty is the determination of the signal 
efficiency. To estimate this effect, we perform two additional fits with 
a modified efficiency function by varying the momentum dependence of 
the $\pi^0$ reconstruction efficiency.
We also perform a fit with a modified detector resolution function:
the resolutions are varied from $4$ to $8\,\mevcc$ instead of the nominal 
$6\,\mevcc$ to take into account the effect of different momentum 
resolutions in MC  and data.
We use different sideband sub-samples to determine the background PDF parameters:
the low-mass sideband only, or the high-mass sideband, or 
$\upns\to\ee$ events only, or $\upns\to\mumu$ events only.
We also vary the signal to background ratio within its errors. 
We considered the effect of the uncertainty of the c.m. energy 
(conservatively taken as $\pm 3$~MeV).

The contribution of all experimental effects to the degradation of $\dtl$ from the 
simultaneous fit of the $\up(2,3S)\pnpn$ sample is smaller than $4.4$.
The corresponding limit for the model uncertainties is $4.5$. 
We combine these two values in quadrature and decrease $\dtl$ from the 
simultaneous fit by $6.3$ in calculations of the $\zbnf$ significance. 
As a result, the $\zbnf$ significance is $6.5\sigma$.
Fits with the $\zbnf$ mass as a free parameter yield values from
$10606$ to $10613\,\mevcc$. We use $\pm 4\,\mevcc$ as a model uncertainty for 
the $\zbnf$ mass.

\begin{table*}
\caption{Systematic uncertainty on the fractions of 
individual channels in the $\upns\pnpn$ final states.}
\medskip
\label{tab:fracsyst}
\begin{tabular*}{\textwidth}{@{\extracolsep{\fill}}lcccccc}\hline\hline
Uncertainty, \%   & Model & Efficiency & Resolution & Background & Beam energy & Sum\\
\hline
$\up(1S)$, $\zbnf$ & $^{+0.5}_{-0.3}$ & $^{+0.2}_{-0.1}$ & $\pm 0.04$ & $\pm 0.07$ & $\pm 0.04$ & $^{+0.5}_{-0.3}$\\
$\up(1S)$, $\zbns$ & $^{+0.5}_{-0.3}$ & $^{+0.2}_{-0.1}$ &  $\pm 0.02$ & $^{+0.13}_{-0.06}$ & $\pm 0.01$ & $^{+0.5}_{-0.3}$\\
$\up(1S)$, $f_2(1275)$&$^{+7.7}_{-4.4}$ & $^{+0.7}_{-0.8}$ &  $\pm 0.02$ & $^{+0.5}_{-0.9}$ & $\pm 0.1$ & $^{+7.8}_{-4.5}$\\
$\up(1S)$, S-wave &$^{+5.5}_{-2.8}$ & $^{+0.6}_{-1.0}$ &  $\pm 0.05$ & $^{+0.9}_{-1.4}$ & $\pm 0.7$ & $^{+5.6}_{-3.4}$\\
\hline
$\up(2S)$, sol. A, $\zbnf$ & $^{+1.4}_{-3.0}$ & $^{+0.6}_{-0.3}$ &  $\pm 2.1$ & $^{+1.8}_{-2.4}$ & $\pm 0.2$ & $^{+3.2}_{-4.4}$\\
$\up(2S)$, sol. A, $\zbns$ & $^{+1.1}_{-0.6}$ & $^{+0.1}_{-0.03}$ &  $\pm 0.8$ & $^{+0.5}_{-0.6}$ & $\pm 0.1$ & $^{+1.5}_{-1.2}$\\
$\up(2S)$, sol. A, $f_2(1275)$&$^{+0.3}_{-0.8}$ & $\pm  0.8$ &  $\pm 0.8$ & $^{+3.6}_{-1.6}$ & $\pm 0.1$ & $^{+3.8}_{-2.1}$\\
$\up(2S)$, sol. A, S-wave &$^{+3.8}_{-0.7}$ & $^{+2.5}_{-2.3}$ &  $\pm 0.5$ & $^{+3.9}_{-6.0}$ & $\pm 0.5$ & $^{+6.0}_{-6.5}$\\
\hline
$\up(2S)$, sol. B, $\zbnf$ & $^{+4.0}_{-11}$ & $^{+0.7}_{-1.6}$ &  $\pm 0.6$ & $^{+0.5}_{-2.2}$ & $\pm 0.6$ & $^{+4.2}_{-11}$\\
$\up(2S)$, sol. B, $\zbns$ & $^{+0.3}_{-0.1}$ & $^{+0.07}_{-0.1}$ &  $\pm 1.0$ & $^{+0.4}_{-0.6}$ & $\pm 0.3$ & $\pm 1.2$\\
$\up(2S)$, sol. B, $f_2(1275)$&$^{+0.4}_{-3.6}$ & $^{+0.8}_{-0.6}$ &  $\pm 0.3$ & $^{+3.2}_{-1.6}$ & $\pm 2.1$ & $^{+3.9}_{-4.5}$\\
$\up(2S)$, sol. B, S-wave &$^{+5}_{-15}$ & $^{+1.5}_{-1.4}$ &  $\pm 0.5$ & $^{+2}_{-10}$ & $\pm 2$ & $^{+6}_{-18}$\\
\hline
$\up(3S)$, $\zbnf$ & $^{+2}_{-5}$  & $\pm 5$ &  $^{+1.3}_{-0.4}$ & $^{+13}_{-8}$ & $\pm 0.8$ & $^{+14}_{-11}$\\
$\up(3S)$, $\zbns$ & $^{+2.7}_{-0.8}$& $^{+1.4}_{-0.6}$ &  $^{+1.4}_{-1.0}$ & $^{+1.1}_{-1.2}$ & $\pm 0.02$ & $^{+3.5}_{-1.9}$\\
$\up(3S)$, S-wave  & $^{+12}_{-7}$  & $\pm 1$ &  $^{+2}_{-5}$ & $^{+13}_{-15}$ & $\pm 0.3$ & $^{+18}_{-17}$\\
\hline\hline
\end{tabular*}
\end{table*}

\section{Conclusion}
We report the observation of $\upf\to\upns\pnpn$ decays with $n=1,2$ and 3.
The measured cross sections,
$\sigma(\ee\to\upf\to\up(1S)\pnpn)=(1.16\pm 0.06\pm 0.10)\, {\rm pb}$,
$\sigma(\ee\to\upf\to\up(2S)\pnpn)=(1.87\pm 0.11\pm 0.23)\, {\rm pb}$, and
$\sigma(\ee\to\upf\to\up(3S)\pnpn)=(0.98\pm 0.24\pm 0.19)\, {\rm pb}$,
are consistent with the expectations from isospin 
conservation based on $\sigma(\upf\to\upns\pipi)$~\cite{br_ypipi,y5s_3body}.

The first observation of a neutral resonance decaying to $\up(2,3S)\pi^0$, 
the $\zbnf$, has been obtained in a Dalitz analysis of 
$\upf\to\up(2,3S)\pnpn$ decays.
The statistical significance of the $\zbnf$ signal is $6.8\sigma$ 
($6.5\sigma$ including experimental and model uncertainties).
Its measured mass, $m(\zbnf)=(10609\pm 4\pm 4)\,\mevcc$, is consistent 
with the mass of the corresponding charged state, the $\zbc(10610)$.
The $\zbns$ signal is not significant in any of the $\up(1,2,3S)\pnpn$ 
channels. 
Our data are consistent with the existence of $\zbns$, but the 
available statistics are insufficient for the observation of this state.

\section*{Acknowledgments}
We thank the KEKB group for the excellent operation of the
accelerator; the KEK cryogenics group for the efficient
operation of the solenoid; and the KEK computer group,
the National Institute of Informatics, and the 
PNNL/EMSL computing group for valuable computing
and SINET4 network support.  We acknowledge support from
the Ministry of Education, Culture, Sports, Science, and
Technology (MEXT) of Japan, the Japan Society for the 
Promotion of Science (JSPS), and the Tau-Lepton Physics 
Research Center of Nagoya University; 
the Australian Research Council and the Australian 
Department of Industry, Innovation, Science and Research;
Austrian Science Fund under Grant No. P 22742-N16;
the National Natural Science Foundation of China under
contract No.~10575109, 10775142, 10875115 and 10825524; 
the Ministry of Education, Youth and Sports of the Czech 
Republic under contract No.~MSM0021620859;
the Carl Zeiss Foundation, the Deutsche Forschungsgemeinschaft
and the VolkswagenStiftung;
the Department of Science and Technology of India; 
the Istituto Nazionale di Fisica Nucleare of Italy; 
The BK21 and WCU program of the Ministry Education Science and
Technology, National Research Foundation of Korea Grant No.\ 
2010-0021174, 2011-0029457, 2012-0008143, 2012R1A1A2008330,
BRL program under NRF Grant No. KRF-2011-0020333,
and GSDC of the Korea Institute of Science and Technology Information;
the Polish Ministry of Science and Higher Education and 
the National Science Center;
the Ministry of Education and Science of the Russian
Federation and the Russian Federal Agency for Atomic Energy;
the Slovenian Research Agency;
the Basque Foundation for Science (IKERBASQUE) and the UPV/EHU under 
program UFI 11/55;
the Swiss National Science Foundation; the National Science Council
and the Ministry of Education of Taiwan; and the U.S.\
Department of Energy and the National Science Foundation.
This work is supported by a Grant-in-Aid from MEXT for 
Science Research in a Priority Area (``New Development of 
Flavor Physics''), and from JSPS for Creative Scientific 
Research (``Evolution of Tau-lepton Physics'').
This work is partially supported by grants of the Russian Foundation for Basic 
Research 12-02-00862, 12-02-01296, 12-02-01032, 12-02-33015 and by the 
grant of the Russian Federation government 11.G34.31.0047.


\begin{thebibliography}{99}

\bibitem{zb_paper}
A. Bondar, A. Garmash, R. Mizuk, D. Santel, K. Kinoshita {\it et al.} (Belle Collaboration),
Phys. Rev. Lett.  {\bf 108}, 122001 (2012).

\bibitem{hb_paper}I. Adachi {\it et al.} (Belle Collaboration),
Phys. Rev. Lett. {\bf 108}, 032001 (2012).

\bibitem{zb_helicity}
I. Adachi {\it et al.} (Belle Collaboration), arXiv:1105.4583 [hep-ex].

\bibitem{danilkin}
I. V. Danilkin, V. D. Orlovsky and Yu. A. Simonov, 
Phys. Rev. {\bf D 85}, 034012 (2012).

\bibitem{bugg}
D. Bugg, Europhys. Lett. {\bf 96}, 11002 (2011).

\bibitem{karliner}
M. Karliner and H. J. Lipkin, arXiv:0802.0649 [hep-ph].

\bibitem{zb_molecular}
A.E. Bondar, A. Garmash, A.I. Milstein, R. Mizuk and M.B. Voloshin,
Phys. Rev. {\bf D 84} , 054010 (2011).

\bibitem{y5s_3body}
I.~Adachi {\it et al.} (Belle Collaboration),
arXiv:1209.6450 [hep-ex].

\bibitem{belle}
A.~Abashian {\it et al.} (Belle Collaboration), Nucl. Instrum. Methods 
 Phys. Res. Sect. {\bf A 479}, 117 (2002); also see detector section in
 J.~Brodzicka {\it et al.}, Prog. Theor. Exp. Phys. (2012) 04D001. 

\bibitem{kekb}
S.~Kurokawa and E.~Kikutani, Nucl. Instrum. Methods Phys. Res. Sect.
 {\bf A 499}, 1 (2003), and other papers included in this Volume;
 T.~Abe {\it et al.}, Prog. Theor. Exp. Phys. (2013) 03A001 and following
 articles up to 03A011.

\bibitem{muid}
A.Abashian {\it et al.}, 
Nucl. Instrum. Meth. {\bf A 491}, 69 (2002).

\bibitem{elid}
K.Hanagaki {\it et al.}, 
Nucl. Instrum. Meth. {\bf A 485}, 490 (2002).

\bibitem{br_ypipi}
K.-F. Chen, W.-S. Hou {\it et al.} (Belle Collaboration),
Phys. Rev. {\bf D 82}, 091106(R) (2010).

\bibitem{kuraev}
E.A.Kuraev, V.S.Fadin, Sov. J. Nucl. Phys. {\bf 41}, 466 (1985).
M.Benayoun, S.I. Eidelman, V.N.Ivanchenko, Z.K.Silagadze, 
Mod. Phys. Lett. {\bf A 14}, 2605 (1999).

\bibitem{Note1} Throughout the paper, the first and second errors are
statistical and systematic, respectively. When one error is shown, it includes
only statistical uncertainty. In such cases the systematic error is omitted.

\bibitem{actis}
S. Actis {\it et al.}, 
Eur. Phys. J. {\bf C 66}, 585 (2010).

\bibitem{sigma_y5s}
S. Esen, A. J. Schwartz {\it et al.} (Belle Collaboration),
Phys. Rev. {\bf D 87}, 031101(R) (2013).

\bibitem{flatte}
S. M. Flatte, Phys. Lett. {\bf B 63}, 224, (1976).

\bibitem{kpipi}
A. Garmash {\it et al.} (Belle Collaboration), 
Phys. Rev. Lett. {\bf 96}, 251803 (2006).

\bibitem{PDG}
J. Beringer {\it et al.} (Particle Data Group), 
Phys. Rev. {\bf D 86}, 010001 (2012) and 2013 partial update
for the 2014 edition.

\bibitem{voloshin}
M.B. Voloshin, Prog. Part. Nucl. Phys. {\bf 61}, 455 (2008);
M.B. Voloshin, Phys. Rev. {\bf D 74}, 054022 (2006) and references therein.


\end{thebibliography}
\end{document}